\documentclass[english,aps,prl,preprint,superscriptaddress, group address, amsmath]{revtex4-2}

\usepackage{braket}
\usepackage{fancyhdr,graphicx}
\usepackage[table,xcdraw]{xcolor}
\usepackage[normalem]{ulem}
\useunder{\uline}{\ul}{}
\usepackage[T1]{fontenc}
\usepackage{chemformula} 
\usepackage[utf8x]{inputenc}
\usepackage{gensymb}
\usepackage{times}
\usepackage{color}
\usepackage{amsfonts}
\usepackage{amssymb}
\usepackage[colorlinks,bookmarks=false,citecolor=blue,linkcolor=red,urlcolor=blue]{hyperref}
\input pdfcolor.tex
\usepackage{colortbl,amsthm,amsmath,amssymb,txfonts}
\usepackage{graphicx}
\usepackage{epsfig,color}
\usepackage{verbatim}
\usepackage{dcolumn}
\usepackage[normalem]{ulem}
\usepackage{bm}
\usepackage{xcolor}

\usepackage{orcidlink}

\usepackage[normalem]{ulem}
\usepackage{cancel}

\usepackage{babel}
\begin{document}

\title{Even denominator fractional quantum Hall states in the zeroth Landau level of ABA trilayer graphene}

\author{Tanima Chanda\orcidlink{0009-0003-3516-8890}}
\thanks{These authors contributed equally to this work.}
\affiliation{Department of Physics, Indian Institute of Science, Bangalore 560012, India}

\author{Simrandeep Kaur\orcidlink{0000-0002-1460-0686}}
\thanks{These authors contributed equally to this work.}
\affiliation{Department of Physics, Indian Institute of Science, Bangalore 560012, India}

\author{Harsimran Singh\orcidlink{0009-0009-8560-0518}}
\affiliation{Department of Physics, Indian Institute of Science, Bangalore 560012, India}

\author{Kenji Watanabe}
\affiliation{Research Center for Electronic and Optical Materials, National Institute for Materials Science, 1-1 Namiki, Tsukuba 305-0044, Japan}

\author{Takashi Taniguchi}
\affiliation{Research Center for Materials Nanoarchitectonics, National Institute for Materials Science, 1-1 Namiki, Tsukuba 305-0044, Japan}

\author{Manish Jain\orcidlink{0000-0001-9329-6434}}
\affiliation{Department of Physics, Indian Institute of Science, Bangalore 560012, India}

\author{Udit Khanna\orcidlink{0000-0002-3664-4305}}
\affiliation{Theoretical Physics Division, Physical Research Laboratory, Navrangpura, Ahmedabad 380009, India}

\author{Ajit C. Balram\orcidlink{0000-0002-8087-6015}}
\affiliation{Institute of Mathematical Sciences, CIT Campus, Chennai 600113, India}
\affiliation{Homi Bhabha National Institute, Training School Complex, Anushaktinagar, Mumbai 400094, India.}

\author{Aveek Bid\orcidlink{0000-0002-2378-7980}}
\email{aveek@iisc.ac.in}
\affiliation{Department of Physics, Indian Institute of Science, Bangalore 560012, India}

\begin{abstract}
Even-denominator fractional quantum Hall states (FQHSs) at half filling are of particular interest because they can host non-Abelian quasiparticles. Here we report the emergence of such states in the zeroth Landau level ($N=0$) of ABA trilayer graphene (TLG), challenging the conventional expectation that they are confined to the first excited Landau level. We observe robust incompressible states at $\nu=7/2$, $9/2$, and $5/2$ with their associated Levin--Halperin daughter states: $\nu=59/17$ and $46/13$ near $7/2$; $\nu=58/13$ and $77/17$ near $9/2$; and $\nu=43/17$ near $5/2$. These states appear exclusively within a finite displacement-field window coincident with crossings between symmetry-broken $N=0$ Landau levels carrying distinct isospin indices. The quantitative correspondence between the calculated crossing loci and the experimentally determined stability regions identifies Landau-level mixing as the microscopic origin. We attribute the stabilization of these even-denominator states to inversion-symmetry breaking in TLG, which enhances valley-resolved Landau-level hybridization and renormalizes short-range Coulomb interactions. Our results expand the landscape of even-denominator FQHSs to multilayer graphene and establish TLG as a tunable platform for realizing non-Abelian anyons.
\end{abstract}

\maketitle

\textit{Introduction} -- The fractional quantum Hall effect (FQHE)~\cite{PhysRevLett.50.1395, PhysRevLett.48.1559} exemplifies strongly correlated topological phases in two-dimensional electron systems. Even-denominator FQHSs are of particular interest because they can host quasiparticles with non-Abelian statistics~\cite{MOORE1991362, PhysRevLett.94.166802, RevModPhys.80.1083}. The canonical example is the $\nu = 5/2$ state~\cite{PhysRevLett.59.1776, 10.1093/nsr/nwu071, Banerjee2018, doi:10.1126/science.abm6571} in GaAs quantum wells~\cite{PhysRevLett.83.3530, PhysRevLett.88.076801}, arising at half-filling of the second Landau level (LL) ($N=1$)~\cite{PhysRevLett.59.1776}, where a node in the orbital wavefunction softens short-range Coulomb repulsion and enables composite-fermion pairing~\cite{annurev:/content/journals/10.1146/annurev-conmatphys-031214-014606, jain2007composite}. In contrast, the lowest LL ($N=0$) is dominated by strong short-range repulsion and stabilizes a compressible composite-fermion Fermi sea at half-filling~\cite{PhysRevB.47.7312}, so even-denominator FQHSs are not expected within conventional theory.

This constraint can be overcome by finite layer thickness and Landau-level (LL) mixing. Finite thickness in wide GaAs wells softens intra-LL interactions and stabilizes $\nu = 1/2$ and $1/4$ states~\cite{ywpx-qm7d, PhysRevB.47.4394, PhysRevLett.101.266804,PhysRevB.103.155306, Singh2024, PhysRevLett.72.3405, PhysRevB.109.035306, PhysRevLett.131.266502}, while strong LL mixing in GaAs, graphene, and 2D hole systems renormalizes the effective interaction and can stabilize paired phases~\cite{PhysRevB.87.245129, PhysRevLett.130.186302, PhysRevLett.129.156801}. Even-denominator states have recently been reported in graphene systems~\cite{Kumar2025, Chen2024, Zibrov17, PhysRevLett.121.226801, Li2017EvenDenominatorBLG, Ki2014, Zibrov16}  indicating such interaction renormalization, although the microscopic mechanism remains unclear.

%

At half-filling, candidate states fall into two broad classes. The first comprises single-component paired states---the non-Abelian Moore--Read Pfaffian~\cite{PhysRevLett.99.236806} and its hole-conjugate anti-Pfaffian~\cite{MOORE1991362, PhysRevLett.117.096802}---in which pairing occurs within a single spin, valley, or other isospin component. The second is the Abelian two-component Halperin-$331$ state~\cite{PhysRevB.82.235312, PhysRevB.110.165402}, which involves correlations between two distinct spin, valley, or isospin components.

Here we demonstrate a distinct mechanism for stabilizing single-component even-denominator pairing in the $N=0$ LL. We observe robust FQHSs at $\nu = 9/2$, $7/2$, and $5/2$ in the monolayer-like band of ABA trilayer graphene (ABA-TLG)~\cite{nm8b-5vgm, PhysRevB.111.235118}, where a compressible composite-fermion liquid is expected. These states emerge exclusively within a narrow displacement-field window when two symmetry-broken $N=0$ LLs with different isospin character approach and cross in energy. We further resolve Levin–Halperin daughter states~\cite{PhysRevB.79.205301, MOORE1991362, Huang21, Kumar2025, Nakamura2020}, placing the paired phases within the Moore–Read universality class.

These results identify ABA-TLG as a platform where controlled LL proximity within the $N=0$ manifold provides a previously unexplored route to stabilizing even-denominator pairing. Unlike GaAs, where $\nu = 5/2$ arises in $N=1$, all even-denominator states observed here occur entirely within the $N=0$ monolayer-like sector. Although even-denominator states at $\nu = \pm 1/2$ in the $N = 0$ LL of monolayer graphene (MLG) have been reported previously~\cite{PhysRevLett.121.226801, Zibrov17}, their daughter hierarchies were not examined. Our results, therefore, not only uncover a distinct stabilization mechanism but also provide compelling evidence for paired-phase character.

Our findings contrast with bilayer graphene (BLG), where incompressible half-filled states are absent despite the convergence of LLs with different isospins~\cite{Huang21, Huang25}. In BLG, inversion symmetry at $D=0$ enforces valley degeneracy of the $N=0$ levels. Contrarily, the lack of intrinsic inversion symmetry in ABA-TLG enables lattice-scale couplings that modify the crossing topology, enhance LL mixing, and stabilize even-denominator FQHSs (see End Matter).

\textit{Results --} Dual graphite-gated hexagonal boron nitride (hBN)-encapsulated ABA-TLG devices were fabricated using a standard dry transfer method~\cite{doi:10.1126/science.1244358, Kaur2024, Pizzocchero2016, mohitNatcomm}. Independent top and bottom graphite gates enable simultaneous control of carrier density $n$ and perpendicular displacement field $D$ (Sec.~1 of Supplemental Material (SM))~\footnote{The Supplemental Material includes details of device fabrication and data at $\nu=5/2$, $\nu=7/2$, and $\nu=9/2$.}. Measurements were performed at $T = 20$ mK unless otherwise noted. We focus primarily on $\nu = 7/2$ and $\nu = 9/2$; $\nu = 5/2$ data are presented in Sec.~S5 of SM~\cite{Note1}.

\begin{figure}[t]
		\includegraphics[width=0.75\columnwidth]{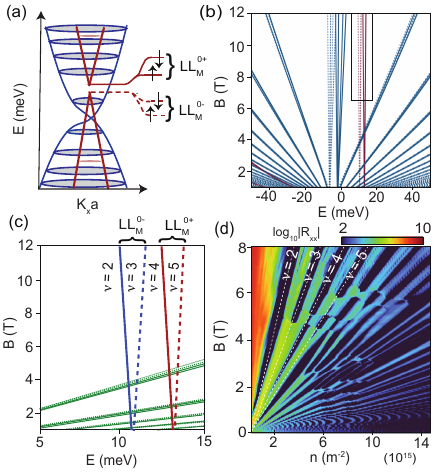}
		\small{\caption{\textbf{Electronic structure and LLs of ABA-TLG.} (a) Schematic band structure of ABA-TLG at $D=0$, showing coexisting monolayer- and bilayer-like bands. The four zeroth LLs of the monolayer-like band ($LL_M^{0}$) are indicated. (b) Calculated LL spectrum at $D=0$. Red (teal) lines denote monolayer-like (bilayer-like) levels; solid and dotted lines correspond to the $K$ and $K'$ valleys, respectively. The black rectangle highlights the spin- and valley-resolved $LL_M^{0}$ manifold hosting the even-denominator FQHSs. (c) Zoomed view of the monolayer-like zeroth LLs. Red (blue) curves denote the $K$ ($K'$) valley, while solid (dashed) lines indicate spin-up (spin-down); green lines mark bilayer-like LLs. (d) Measured $R_{xx}$ in the $B-n$ plane, identifying the filling-factor sequence of the monolayer-like sector.
}  \label{fig: Fig1}}
	\end{figure}

ABA-stacked (Bernal-stacked) trilayer graphene comprises three graphene layers, with the middle layer shifted relative to the aligned top and bottom layers by one carbon–carbon bond vector (Fig.~S13 of SM~\cite{Note1}). At $D=0$, ABA-TLG hosts coexisting monolayer- and bilayer-like bands protected by mirror symmetry (Fig.~\ref{fig: Fig1}(a)), with their Dirac points offset in energy~\cite{PhysRevB.87.115422, PhysRevLett.121.167601}. 
The calculated LL spectrum (Fig.~\ref{fig: Fig1}(b)) shows monolayer-like (red) and bilayer-like (blue) LLs. We label monolayer-like LLs as $LL^{\beta\gamma}_{M,s}$, where $\beta$ is the orbital index, $\gamma=\pm$ the valley, and $s=\uparrow,\downarrow$ the spin. For $B \gtrsim 6$ T, $\nu = 2,3$ arise from spin-split $LL^{0-}_{M}$ levels, while $\nu = 4,5$ originate from $LL^{0+}_{M}$ (Fig.~\ref{fig: Fig1}(c)), consistent with $R_{xx}(n,B)$ (Fig.~\ref{fig: Fig1}(d)).

\begin{figure*}[t!]
\includegraphics[width=\textwidth]{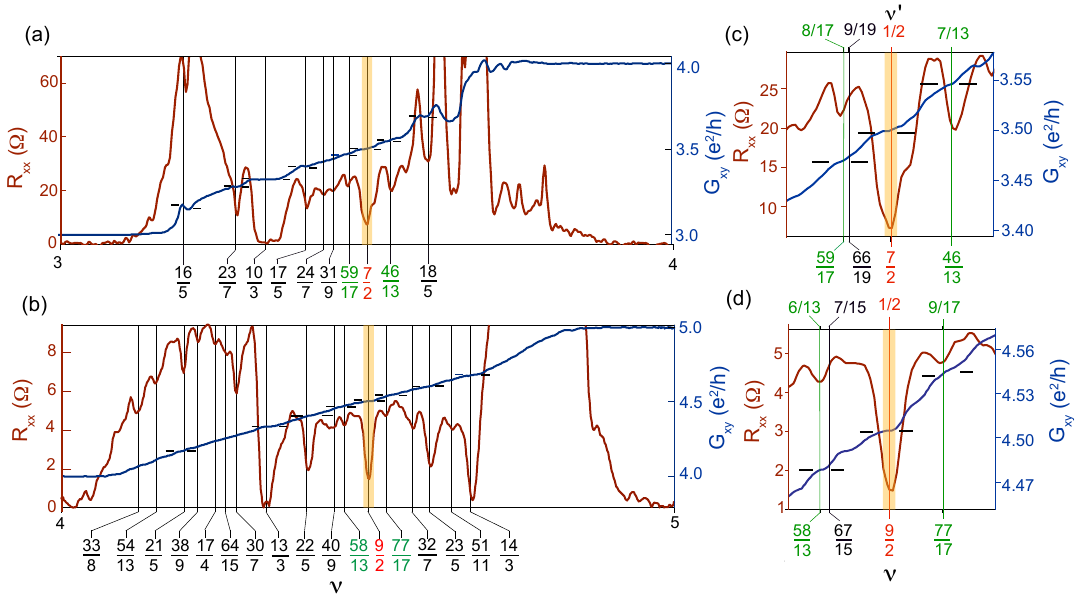}
  \small{\caption{\textbf{Even-denominator FQHSs.} 
Longitudinal resistance $R_{xx}$ (left axis, brown) and transverse conductance $G_{xy}$ (right axis, blue) versus filling factor $\nu$ measured at $T=20$~mK. 
(a) $\nu=3$–$4$ at $B = 15.7$~T, $D = -0.079$~V/nm (device 1).
(b) $\nu=4$–$5$ at $B = 16$~T, $D = 0.167$~V/nm (device 2).
(c,d) Zoomed-in views highlighting the even-denominator states and their Levin–Halperin daughter states. The upper axis shows the effective filling factor $\nu' =  \nu-N_L$, where $N_L$ is the lowest filled LL. Orange vertical shade mark even-denominator states; green vertical lines indicate observed Levin–Halperin daughters; black vertical lines denote the locations of daughter states expected for a Halperin-331 and Halperin-113 phase but not observed. Black horizontal lines are guides to the eye, indicating the expected quantized Hall conductance values.  \label{fig:Fig2}}}
\end{figure*}

  	\begin{figure}[h!]
		\includegraphics[width=0.75\columnwidth]{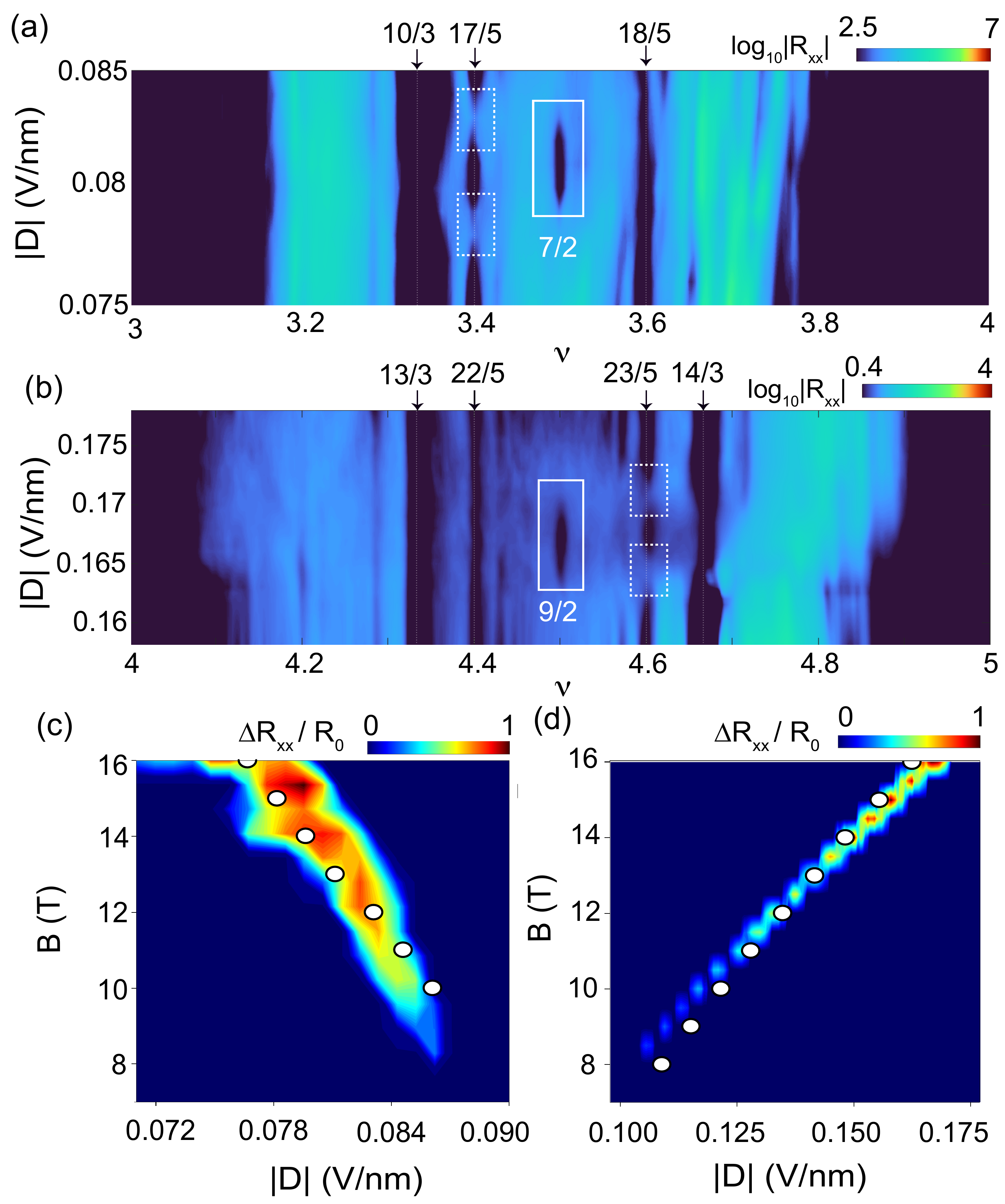}
		\small{\caption{\textbf{Evolution of even denominator FQHSs with $D$ and $B$.} $R_{xx}$ as a function of $\nu$ and $|D|$ near (a) $\nu = 7/2$ ($B = 12$ T) (device 1) and (b) $\nu = 9/2$ ($B = 16$ T) (device 2). White rectangles highlight even-denominator states. {White dashed rectangles indicate the regions of suppressed odd-denominator states.} Normalized suppression $\Delta R_{xx}/R_0$ in the $B-D$ plane at (c) $\nu = 7/2$ and (d) $\nu = 9/2$. White circles denote calculated LL crossing points between (c) $LL^{0+}_{M,\uparrow}$ and $LL^{0-}_{M,\downarrow}$ LLs and (d) $LL^{0+}_{M,\downarrow}$ and $LL^{0-}_{M,\uparrow}$ LLs. The correspondence between the minima of the even-denominator states and the LL crossings demonstrates quantitative correspondence between LL proximity and stability of the even-denominator phases.
        \label{fig:fig3}}}
	\end{figure}
    

Figs.~\ref{fig:Fig2}(a,b) show longitudinal resistance $R_{xx}$ and transverse conductance $G_{xy}$ over the filling-factor ranges $\nu = 3$--$4$ and $4$--$5$ at $B = 15.7$~T and $16$~T, respectively, with finite $D$. In addition to the odd-denominator Jain sequence, pronounced incompressible states appear at $\nu = 7/2$ and $9/2$, evidenced by deep $R_{xx}$ minima and half-integer $G_{xy}$ plateaus. The $\nu = 7/2$ state is accompanied by Levin--Halperin daughter states at $\nu = 59/17~(3+8/17)$ and $46/13~(3+7/13)$, consistent with a Moore--Read Pfaffian phase~\cite{PhysRevB.79.205301, PhysRevB.110.165402}. By contrast, the $\nu = 66/19~(3+9/19)$ daughter expected for a Halperin--331 state is absent, although $\nu = 46/13~(3+7/13)$ is a daughter state filling common to both Moore--Read Pfaffian and Halperin--331 state (Fig.~\ref{fig:Fig2}(c)). Conversely, the $\nu = 9/2$ state is flanked by daughter states at $\nu = 58/13~(4+6/13)$ and $77/17~(4+9/17)$, consistent with an anti-Pfaffian phase. We do not observe the $\nu = 67/15~(4+7/15)$ daughter expected for a Halperin--113 state, although $\nu = 77/17~(4+9/17)$ is daughter state fraction common to both anti-Pfaffian and Halperin--113 state (Fig.~\ref{fig:Fig2}(d)).

\begin{figure*}[t]
		\includegraphics[width=\textwidth]{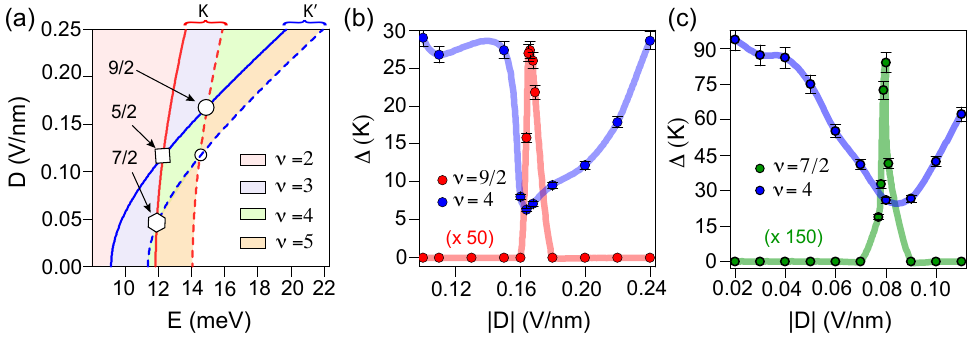}
		\small{\caption{\textbf{LL crossings and the origin of even denominator FQHSs.} (a) Calculated energy spectra of zeroth monolayer-like LLs versus $D$ at $B=16$~T. Red (blue) curves correspond to $K$ ($K'$) valleys; solid (dashed) lines denote spin-up (spin-down). Symbols mark crossings between spin- and valley-resolved monolayer-like $LL^{0\pm}_{M,s}$ Landau levels associated with the observed even-denominator states. (b-c) Activation gaps as a function of $|D|$, comparing half-filled states with adjacent integer states for (b) $\nu = 9/2$ and $\nu =4$ measured at $B=16$~T (device 2), and (c) $\nu = 7/2$ and $\nu =4$ measured at $B=13$~T (device 1). The gap for the half-filled states peaks near the LL crossing, while the corresponding integer gap is suppressed, evidencing LL mixing near degeneracy. {The lines in the plots are guides to the eye.}
\label{fig:fig4}}}
	\end{figure*}
    
A key observation is that even-denominator states appear only within a finite window of the displacement field. Figs.~\ref{fig:fig3}(a,b) show $R_{xx}(\nu,D)$ near $\nu = 7/2$ ($B=12$ T) and $\nu = 9/2$ ($B=16$ T); outside this window the states revert to compressible behavior, with daughter states exhibiting similar $D$-dependence (Sec.~S3.2 and S4.2).

The onset of $\nu = 7/2$ is accompanied by suppression of nearby odd-denominator states:   $\nu=17/5$ weakens as $\nu=7/2$ develops and re-emerges after stabilization ({marked by white dashed rectangles in} Fig.~\ref{fig:fig3}(a)). Similarly, $\nu=23/5$ is suppressed at two $D$-regions as $\nu=9/2$ stabilizes ({marked by white dashed rectangles in} Fig.~\ref{fig:fig3}(b)). Notably, fractional states farther from half-filling do not exhibit comparable suppression (Sec.~S3.1 and Sec.~S4.1 of SM~\cite{Note1}). These correlated changes demonstrate that the Landau-level crossings strongly influence the nearby fractional quantum Hall states. A detailed understanding of the evolution of the odd-denominator states requires further study and is beyond the scope of the present work. 
The stability regions of the even-denominator FQHSs are quantified in Figs.~\ref{fig:fig3}(c-d) by the normalized suppression $\Delta R_{xx}/R_0$ at $\nu = 7/2$ and $\nu =9/2$ in the $B$–$D$ plane (SM Sec.~S2~\cite{Note1}). The even-denominator states trace well-defined ridges in this parameter space, demonstrating correlated tuning of their stability by the magnetic and displacement fields.

To elucidate the microscopic origin of this behavior, we compute the $D$-dependent LL spectrum using a tight-binding model~\cite{PhysRevB.87.115422, PhysRevLett.121.167601} (Sec.~S7~\cite{Note1}). At $B=16$~T,  finite $D$ leads to crossings between the four spin-resolved zeroth-LL of the monolayer-like band (Fig.~\ref{fig:fig4}(a)). As $D$ increases, the levels $LL^{0+}_{M,\uparrow}$ (red solid line) and $LL^{0-}_{M,\downarrow}$ (blue dashed line) approach and intersect near $D\sim 0.05$~V/nm at $B=16$~T (white filled hexagon in Fig.~\ref{fig:fig4}(a)). This crossing coincides with the experimentally observed stability window of the $\nu=7/2$ state. The calculated crossing of $LL^{0+}_{M,\uparrow}$ and $LL^{0-}_{M,\downarrow}$ LLs in the $B$–$D$ plane (white circles in Fig.~\ref{fig:fig3}(c)) coincide with the experimentally determined stability region. 

Similarly, the crossing between $LL^{0+}_{M,\downarrow}$ and $LL^{0-}_{M,\uparrow}$ (indicated by the white filled circle in Fig.~\ref{fig:fig4}(a)) aligns with the $\nu=9/2$ stability region. The two states exhibit opposite evolution with $D$: $\nu=7/2$ shifts to lower $B$, while $\nu=9/2$ shifts to higher $B$ (Fig.~\ref{fig:fig3}(c, d)). This follows from Zeeman-driven evolution of spin-split LLs: increasing $B$  shifts  the crossing between $LL^{0-}_{M,\uparrow}$ and $LL^{0+}_{M, \downarrow}$ to higher $D$, while that between $LL^{0-}_{M, \downarrow}$ and $LL^{0+}_{M,\uparrow}$ moves toward lower $D$. This systematic evolution strongly supports a crossing-driven mechanism rather than disorder-induced effects. 

In Sec.~S5 of the SM~\cite{Note1}, we report a $\nu=5/2$ state accompanied by $\nu=43/17$ (which is a common daughter of both the Halperin--113 and anti-Pfaffian states), appearing near $D\sim 0.097$ V/nm and $B=12$~T. As shown in Fig.~\ref{fig:fig4}(a), this region is close to a same-spin crossing between $LL^{0+}_{M,\uparrow}$ (red solid line) and $LL^{0-}_{M, \uparrow}$ (blue solid line). Although this same-spin crossing occurs near both $\nu=5/2$ and its particle–hole conjugate $\nu=7/2$, an even-denominator state appears only at $\nu=5/2$, demonstrating that particle–hole symmetry is not exact within the $N=0$ manifold of ABA-TLG. LL mixing and displacement-field–induced valley hybridization break the ideal conjugation symmetry, rendering pairing energetically selective. Although each valley-resolved Landau-level crossing is adjacent to two particle-hole-related half-filled states, our data show that at most one develops an incompressible even-denominator state, demonstrating that Landau-level proximity alone is insufficient to determine the resulting paired phase.

Not all crossings yield pairing: no even-denominator state is observed near the crossing between $LL^{0-}_{M,\downarrow}$ (blue dashed line) and $LL^{0+}_{M,\downarrow}$ (red dashed line) (marked by the empty circle in Fig.~\ref{fig:fig4}(a)), where the system remains compressible. This shows that LL proximity alone is insufficient; stabilization depends on internal quantum numbers, wavefunction structure (valley/layer/sublattice structure at finite $D$), the interaction matrix elements generated by lattice-scale symmetry breaking, and the effective detuning from the crossing.


The even-denominator states emerge only within a finite displacement-field window around the Landau-level crossing. In the absence of disorder, the paired phase is still expected to remain stable over a finite range of $D$, provided that the interaction energy gained from pairing exceeds the detuning energy associated with the level splitting. The observed stability window, therefore, reflects the combined effects of Landau-level proximity, interaction-driven pairing, and inversion-symmetry-breaking-induced valley hybridization. Disorder broadening can further enhance the overlap between the crossing levels, increasing the range over which mixing occurs. Using the spectral width $\Gamma \approx 4$ K extracted from Shubnikov--de Haas oscillations~\cite{frpm-3fs8}, we estimate an additional overlap scale $\Delta D_{\rm dis} \sim \Gamma/(\partial E/\partial D)$, which is comparable to the experimentally observed width (Sec.~S4.3 of SM~\cite{Note1}).

Activation-gap measurements show anti-correlation between integer and even-denominator gaps. In the $D$--range where the $\nu=9/2$ gap peaks, the gap defining the $\nu=4$ integer state---between $LL^{0-}_{M,\uparrow}$ and $LL^{0+}_{M,\downarrow}$---is strongly suppressed (Fig.~\ref{fig:fig4}(b)). A similar anti-correlation is observed between the $\nu=7/2$ state and its neighboring integer gap (Fig.~\ref{fig:fig4}(c); Sec.~S4 of the SM~\cite{Note1}). These observations indicate that enhanced LL mixing near crossings renormalizes interactions and promotes the formation of even-denominator states. Notably, all observed even-denominator states originate within the spin- and valley-resolved monolayer-like $N=0$ sector, while the bilayer-like LLs remain energetically separated and fully occupied over the relevant filling-factor range.

\textit{Discussion -- } Recent experiments have reported even-denominator FQHSs at $\nu = -3/2, 3/2, -9/2,$ and $9/2$ in ABA-TLG~\cite{Chen2024}, with the first three occurring in the $N=1$ orbital and $\nu=9/2$ in $N=0$. Here we establish a distinct mechanism for stabilizing even-denominator states entirely within the monolayer-like $N=0$ manifold. We observe $\nu = 9/2, 7/2,$ and $5/2$ in a regime where two symmetry-broken $N=0$ Landau levels approach and cross, demonstrating that displacement-field-tuned level proximity can drive pairing within the lowest orbital.

The even-denominator states are accompanied by suppression of nearby odd-denominator states (Secs.~S3.1, S4.1 of SM~\cite{Note1}), indicating a pairing instability that competes with conventional composite-fermion FQHSs. Similar behavior has been reported in monolayer graphene near $\nu=-1/2$~\cite{Zibrov17}. No comparable effect is observed near $\nu=5/2$, suggesting sensitivity to the crossing topology and spin configuration.

Long-range Coulomb interaction in graphene approximately preserves $SU(4)$ symmetry~\cite{PhysRevB.87.245129}. Consequently, LL mixing typically occurs between different orbitals, while intra-orbital valley mixing is suppressed. In contrast, the states reported here arise from mixing between two valley-resolved $N=0$ LLs. 
At $\nu=7/2$ and $\nu=9/2$, the relevant LLs have opposite spin polarization, whereas at $\nu=5/2$, they share the same spin. Crossings between LLs with opposite real spins typically produce Ising quantum Hall ferromagnetism with resistance spikes and first-order transitions \cite{doi:10.1126/science.290.5496.1546}; no such signatures are observed here. This indicates that pairing is not driven by spin-domain physics but by valley-resolved LL proximity and interaction-enhanced mixing within the $N=0$ manifold.

Even-denominator FQHSs have been reported in MLG and BLG~\cite{Kumar2025, Chen2024, Zibrov17, PhysRevLett.121.226801, PhysRevB.106.155132, Li2017EvenDenominatorBLG, Zibrov16, PhysRevX.12.031019}, but the mechanism in ABA-TLG is qualitatively distinct. In MLG, pairing arises at interaction-driven isospin transitions, while in BLG, inversion symmetry preserves valley equivalence and suppresses intra-orbital mixing. In contrast, ABA-TLG lacks inversion symmetry, and displacement-field-tuned LL crossings enable controlled valley hybridization within the $N=0$ orbital. The resulting inequivalent valley wavefunctions generate additional isospin-dependent interactions absent in inversion-symmetric systems. Although coupling to bilayer-like sectors is allowed at finite $D$, the $N=0$ monolayer-like LLs remain predominantly orbital-pure with only weak higher-orbital admixture, insufficient by itself to stabilize pairing, consistent with the absence of such states in BLG even at large $D$~\cite{Zibrov2017}.

Microscopically, valley-resolved mixing in ABA-TLG originates from lattice-scale corrections to the Coulomb interaction that break valley equivalence. The absence of inversion symmetry makes $|0,+\rangle$ and $|0,-\rangle$ inequivalent and enables valley-dependent interaction renormalization near LL crossings (Sec.~S8 SM~\cite{Note1}). Consistently, crossings between valley-distinct $N=0$ LLs in BLG do not yield incompressible half-filled states~\cite{Zibrov2017, PhysRevX.12.031019}. A finite displacement field further lifts residual valley symmetry, enhancing hybridization and modifying short-range interaction terms, thereby providing a natural route for LL mixing to renormalize interactions within the $N=0$ manifold.

Although lattice-scale interactions are weaker than the Coulomb scale, they are amplified near LL degeneracy. Enhanced mixing renormalizes short-range interactions, destabilizes the composite-fermion Fermi sea, and favors paired even-denominator phases, making LL proximity a tuning parameter for pairing. A microscopic theory incorporating lattice corrections and realistic LL mixing is required for a quantitative understanding.

Even-denominator states in half-filled $N=0$ LLs of MLG have been attributed to two-component physics~\cite{Zibrov17, PhysRevLett.121.226801, PhysRevB.106.155132}. In contrast, inversion-symmetry breaking in ABA-TLG and the observed Levin–Halperin daughter hierarchy indicate that $\nu=7/2$ is consistent with a one-component Pfaffian, while $\nu=9/2$ is consistent with an anti-Pfaffian phase. Although definitive identification requires thermal Hall or edge probes, the phenomenology strongly favors a one-component paired ground state over a two-component Halperin state.

To conclude, we report even-denominator FQHSs and their daughter states in the $N=0$ LL of graphene, establishing a pairing mechanism distinct from $N=1$ physics in GaAs and BLG. Displacement-field-tuned LL proximity and inversion-symmetry breaking in ABA-TLG provide a controllable route to engineering paired quantum Hall states, with future interferometry, thermal Hall, and tunneling spectroscopy~\cite{Hu2025} offering routes to probe non-Abelian anyons.

\begin{acknowledgments}
We acknowledge helpful discussions with David Mross, Dmitri Feldman, and Steven Simon. U.K. acknowledges support from the Department of Space (DOS), Government of India. A. C. B. thanks the Science and Engineering Research Board (SERB) of the Department of Science and Technology (DST) for financial support through the Mathematical Research Impact Centric Support (MATRICS) Grant No. MTR/2023/000002. A.B. acknowledges funding from the Department of Science and Technology, Govt of India (SP/ANRF-24-0117). K.W. and T.T. acknowledge support from the JSPS KAKENHI (Grant Numbers 21H05233 and 23H02052) and World Premier International Research Center Initiative (WPI), MEXT, Japan. 
\end{acknowledgments}


\begin{center}
\textbf{\large End Matter}
\end{center}

In the main text, we conjectured that the lack of inversion symmetry in ABA-TLG introduces additional lattice-scale couplings that promote LL mixing and stabilize the even-denominator FQHS in the regime where two LLs with different valley indices are close in energy. In this section, we discuss why this might be the case.

In an ideal (zero-width and no LL mixing) two-component $N=0$ LL system, where the Coulomb interaction is SU$(2)$ symmetric, and the two components are exactly degenerate (as is the case with real spins at zero Zeeman energy), at half-filling a pseudospin-singlet composite fermion Fermi sea is expected to be the ground state since it has the lowest energy among the various candidate states at $\nu{=}1/2$ (see Table~\ref{tab:one_half}) (Note that for any two-body interaction, like the Coulomb one, the Moore-Read Pfaffian~\cite{PhysRevLett.99.236806} and its hole-conjugate, the anti-Pfaffian~\cite{ Lee07}, have identical energies in the thermodynamic limit). Therefore, with an $SU(2)$ symmetric interaction like the ideal Coulomb in the $N=0$ LL, which is strongly repulsive at short distances, an FQHE state is unlikely to materialize at half-filling, in contrast with our experimental observations. 

However, the interactions' SU$(2)$ symmetry is broken in the valley sector in graphene~\cite{aliceafisher2006}. Lattice-scale corrections in monolayer (and bilayer) graphene reduce the symmetry of the two-body interactions down to $U(1){\times}Z_{2}$~\cite{kharitonov2012mlg}, where the $U(1)$ arises from the conservation of lattice momentum and corresponds to the conservation of the difference in the number of electrons between the two valleys, and the $Z_{2}$, which renders the interactions invariant under the exchange of the valley indices, is a manifestation of the microscopic inversion symmetry in the lattice. Despite this reduced symmetry, the separate charge conservation in each valley [imposed by the $U(1)$] rules out any LL-mixing between the two LLs with different valleys. Therefore, the interaction at and near the crossing point of two $N=0$ LLs with different valleys is also likely to be strongly repulsive at short distances. This expectation is in line with experiments probing the crossing of two LLs with $N=0$ orbital index but different valleys in BLG, where no FQHS is observed at half-filling~\cite{Huang21}. 

Thus, we conjecture that when two LLs approach each other in ABA-TLG, an avoided level crossing occurs. Near the minimal gap point, LL mixing is enhanced. The orbitals of the two bare $N=0$ LLs hybridize in a nontrivial manner, breaking the $U(1)$ symmetry and allowing LL mixing, which modifies the effective electron–electron interaction in the lower LL of interest. Generically, in this two-component system, all three interactions $V^{\uparrow, \uparrow}(r)$, $V^{\uparrow, \downarrow}(r)=V^{\downarrow, \uparrow}(r)$, and $V^{\downarrow, \downarrow}(r)$ [we use the notation for spins with the understanding that these refer to pseudospins with specific valley-spin combinations] will be different since $Z_{2}$ symmetry is absent. The orbitals hybridize in a complex way. In this $SU(2)$-symmetry-broken setting, one possibility is that the two-component Halperin-331 is realized.

\begin{table*}[t]
\begin{ruledtabular}
\begin{tabular}{|l|c|c|c|c|}
{state} &
{per-particle Coulomb energy} &
{Wen-Zee shift, $\mathcal{S}$} &
{chiral central charge, $c_{-}$} &
{excitations} \\ \hline
spin-singlet CFFS        & $-0.46961(1)$~\cite{Balram17} & 2   & --            & Abelian      \\
fully polarized CFFS     & $-0.4656(1)$~\cite{Balram17}  & 2   & --            & Abelian      \\
Moore–Read Pfaffian      & $-0.45682(3)$~\cite{Balram20b} & 3   & $3/2$~\cite{MOORE1991362} & non-Abelian  \\
anti-Pfaffian            & $-0.45682(3)$~\cite{Balram20b} & $-1$ & $-1/2$~\cite{PhysRevLett.99.236806,Lee07} & non-Abelian \\
Halperin–331             & $-0.46325(1)$                 & 3   & 2~\cite{Wen90,Wen92b} & Abelian     \\
\end{tabular}
\end{ruledtabular}
\caption{\label{tab:one_half} Candidate states at half-filling, their per-particle Coulomb energy in the thermodynamic limit in the $n{=}0$ LL, Wen-Zee shift $\mathcal{S}$~\cite{Wen92} on the sphere, chiral central charge $c_{{-}}$ and nature of their excitations (Abelian/non-Abelian) are given. The Hall viscosity $\eta_{H}{=}\hbar \mathcal{S}/(8\pi\ell^{2})$~\cite{Read09}. The thermal Hall conductance $\kappa_{xy}{=}c_{{-}}[\pi^2 k_{\rm B}^2 /(3h)]T$~\cite{Kane97} (filled LLs make an additional integral contribution to $c_{-}$). The thermal Hall conductance of the gapless composite fermion Fermi seas (CFFSs) is not quantized.  }
\end{table*}

The other possibility is that in the lower LL of interest, due to LL mixing with the upper LL, the short-range part of the Coulomb repulsion is softened (analogous to what happens as one goes from the lowest LL to the second LL of GaAs). The Moore-Read Pfaffian or the anti-Pfaffian states can then be stabilized over the fully polarized CF Fermi sea in this lower LL (single-component system) of interest. On one flank of $5/2$, we observe FQHS at $\nu=43/17$, which could be the Levin-Halperin daughter state~\cite{PhysRevB.79.205301} of either the anti-Pfaffian or the Halperin-113~\cite{PhysRevB.110.165402, Zheltonozhskii24}. However, since the Halperin-113 wave function phase separates~\cite{Gail08, Simon25} [although a legitimate FQHS sharing the same topological properties as Halperin-113 could be realized, no evidence for such a scenario has been presented in the literature], we surmise that the $5/2$ state likely resides in the same topological phase as the anti-Pfaffian.

Importantly, $\nu = 59/17$ is a Levin–Halperin daughter of the Moore–Read state, but not of the Halperin-331 state~\cite{PhysRevB.110.165402}, strongly suggesting that the $7/2$ state lies in the Moore–Read phase. Similarly, the $\nu = 58/13$ state is a Levin–Halperin daughter of the Moore–Read state, but not of the Halperin-113 state, pointing to the possibility that the $9/2$ also lies in the Moore–Read phase. Moreover, the FQHSs away from half-filling at the displacement field, where the half-FQHSs are the strongest, fit the standard Jain sequence of one-component composite fermion states. For completeness, we note here that an alternative mechanism for an  FQHS at $8/17$, that does not rely on the Levin-Halperin daughter-state construction, was recently proposed in Ref.~\cite{Balram24a}. Still, it is unclear whether that state can be stabilized on this platform. The scenario for single-component states that we have presented is akin to what has been experimentally observed in the $N{=}1$ LL of BLG~\cite{Huang21, Kumar2025, Zibrov2017} and in the $N{=}0$ LL of wide-quantum wells of GaAs~\cite{Singh2024}. There has also been theoretical work supporting the one-component nature of the FQHSs realized in the settings mentioned in the previous line~\cite{PhysRevB.103.155306, Zhu20a, Zhu16, Balram21b, PhysRevB.109.035306}. Nevertheless, we cannot conclusively rule out two-component FQHSs. The different candidate FQHSs at half-filling can be unambiguously distinguished through thermal Hall measurements (see Table~\ref{tab:one_half}).

\clearpage

\begin{center}
	\textbf{\large Supplementary Material}
\end{center}

\addto{\captionsenglish}{\renewcommand{\refname}{\textbf{Supplementary References}}}
\addto\captionsenglish{\renewcommand{\figurename}{}}
\addto\captionsenglish{\renewcommand{\tablename}{}}
\renewcommand{\thesection}{S\arabic{section}}
\renewcommand{\thesubsection}{S\arabic{section}.\arabic{subsection}}
\renewcommand{\thefigure}{{Figure S}\arabic{figure}}
\renewcommand{\thetable}{{Table S\arabic{table}}}

\setcounter{figure}{0}
\setcounter{equation}{0}
\setcounter{section}{0}
\setcounter{table}{0}

In this Supplementary Material, we provide (i) device fabrication details, (ii) quantitative definition of the normalized resistance metric, (iii) displacement-field evolution of $\nu = 9/2$, $7/2$, and $5/2$ including daughter states, (iv) analysis of disorder-broadened LL crossings, (v) complete $\nu–D$ mapping of LL crossings, (vi) activation gap analysis, and (vii) tight-binding simulations.

\section{Device fabrication and characterization.}

\begin{figure*}[h]
	\includegraphics[width=0.8\columnwidth]{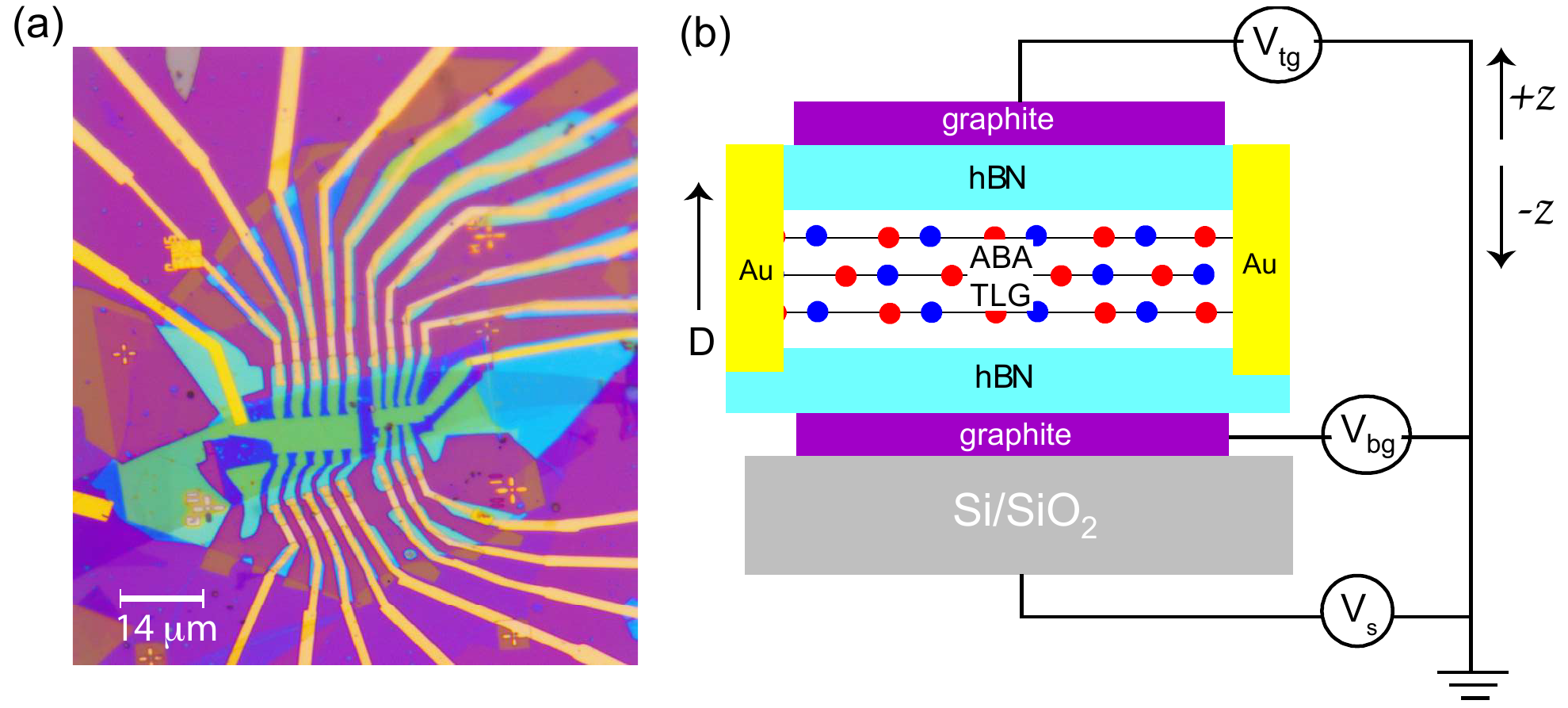}
	\caption{\textbf{Details of Device Schematics.} (a) Optical image of the device. The scale bar is $14~\mu$m. (b) Schematic of the device geometry. The vertical arrow indicates the direction of positive $D$ (along the $+z$ direction).}
	\label{fig:figS1}
\end{figure*}

ABA-TLG, hexagonal boron nitride (hBN), and graphite flakes were mechanically exfoliated onto $280$ nm SiO$_2$/Si substrates. ABA stacking was identified using optical contrast and Raman spectroscopy~\cite{Cong2011, Nguyen2014}, with characteristic splitting and intensity ratios of Raman modes used to distinguish ABA from ABC stacking. The typical lateral size of selected ABA-TLG flakes was $\sim 20-30\mu$m.

The van der Waals heterostructure was assembled using the standard dry pick-up and transfer technique~\cite{Pizzocchero2016, PhysRevLett.129.186802, jat2024}. The final stack consists of graphite/hBN/ABA-TLG/hBN/graphite, enabling dual-gated operation with atomically flat interfaces. One-dimensional edge contacts were defined using electron-beam lithography followed by reactive ion etching in $\mathrm{CHF_3/O_2}$ gas and subsequent Cr/Pd/Au thermal deposition~\cite{doi:10.1126/science.1244358, PhysRevLett.126.096801}. The device was then etched into a Hall bar geometry for longitudinal and Hall transport measurements.

\ref{fig:figS1}(a) shows optical image of device. \ref{fig:figS1}(b) presents a schematic of the device geometry and dual-gate configuration. Measurements were performed in a dilution refrigerator with a base temperature of $20$~mK. 

The typical field-effect mobility extracted from low-field transport is $\mu \sim 620,000$~$\mathrm{cm^2V^{-1}s^{-1}}$ at $T = 20$ mK, with a residual carrier density $n^* \sim  7.81\times10^{9}$~$\mathrm{cm^{-2}}$ for this device. From Shubnikov-de Haas oscillations, we estimate a disorder broadening $\Gamma \sim 4$ K (see Sec.~S4C). These parameters are consistent with the spectral broadening used in the analysis of Landau-level (LL) crossings.


The carrier density $n$ and perpendicular displacement field $D$ were tuned independently using the top and bottom graphite gates according to
\begin{equation}
	D=\frac{C_{bg}V_{bg}-C_{tg}V_{tg}}{2\epsilon_0}, 
	\qquad 
	n=\frac{C_{bg}V_{bg}+C_{tg}V_{tg}}{e},
\end{equation}
where $C_{bg}$ and $C_{tg}$ are the geometric capacitance per unit area of the back and top gates, and $V_{bg}$ and $V_{tg}$ are the corresponding gate voltages.

The capacitance was extracted from the slopes of Landau fan diagrams in the quantum Hall regime and independently cross-checked using SdH oscillation frequencies.  

To prevent the formation of unintended $p$--$n$ junctions near the contacts, the graphene leads extending outside the dual-gated region were electrostatically doped using the global SiO$_2$/Si back gate. The black arrow in \ref{fig:figS1}(b) indicates the direction of the positive displacement field $D$ (along $+z$ direction), corresponding to an electric field pointing from the bottom layer toward the top layer.

In the theoretical modeling, the intralayer electrostatic potential induced by the external displacement field is parametrized by $\Delta_1$. To convert $\Delta_1$ to $D$, we use~\cite{PhysRevLett.132.096301}
\begin{equation}
	\Delta_1 = -\left(\frac{d_{\perp}}{2\epsilon_{TLG}}\right) D\, e,
\end{equation}
which yields
\begin{equation}
	\Delta_1 (\mathrm{meV}) = 85\, D (\mathrm{V/nm}).
\end{equation}

Here, $d_{\perp}=0.67$ nm is the separation between the outer layers of ABA-TLG, $\epsilon_{TLG}=4$ is the effective dielectric constant of TLG, and $e$ is the elementary charge. This conversion assumes uniform electrostatic screening within the TLG and neglects self-consistent Hartree corrections, which are expected to introduce only small quantitative modifications over the displacement-field range explored here.

\section{Definition of ${\Delta R_{xx}}/{R_{0}}$ \label{S2}}

\begin{figure*}[h]
	\includegraphics[width=0.75\columnwidth]{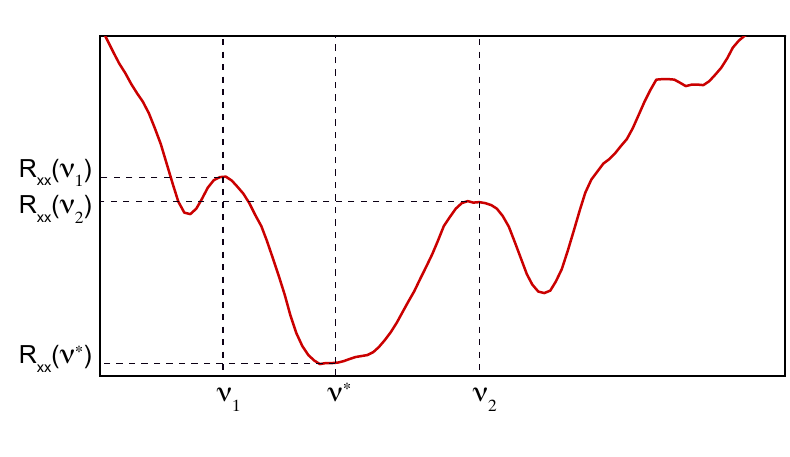}
	\caption{\textbf{Extraction of the normalized resistance dip $\Delta R_{xx}/R_0$.} 
		Representative $R_{xx}(\nu)$ trace illustrating the procedure used to quantify the strength of an incompressible state at filling factor $\nu^\ast$. The minimum resistance $R_{xx}^{\mathrm{min}} = R_{xx}(\nu^\ast)$ is identified at $\nu^\ast$. The resistance at the two nearest local minima at $\nu = (\nu_1, \nu_2)$ are marked as $R_{xx}(\nu_1)$ and $R_{xx}(\nu_2)$. These parameters are used to calculate the normalized resistance dip $\Delta R_{xx}/R_0$ (Sec.\ref{S2}). }
	\label{fig:figS2}
	\label{fig:figS2}
\end{figure*}

To quantify the strength of the FQHSs, we define a normalized resistance dip $\Delta R_{xx}/R_0$ extracted from longitudinal resistance traces $R_{xx}(\nu)$ at fixed magnetic field $B$ and displacement field $D$. For a given filling factor $\nu^\ast$ corresponding to an incompressible state, the minimum longitudinal resistance is denoted by 
$R_{xx}^{\mathrm{min}} \equiv R_{xx}(\nu^\ast)$ (\ref{fig:figS2}).

The reference background resistance $R_0$ is obtained by linearly interpolating between the two nearest local maxima of $R_{xx}$ located at filling factors $\nu_1$ and $\nu_2$ immediately adjacent to $\nu^\ast$. Explicitly,
\begin{equation}
	R_0 = R_{xx}(\nu_1) + 
	\frac{\nu^\ast - \nu_1}{\nu_2 - \nu_1}
	\left[ R_{xx}(\nu_2) - R_{xx}(\nu_1) \right].
\end{equation}

This interpolation removes the slowly varying background contribution and isolates the intrinsic depth of the incompressible minimum. The resistance dip is then defined as
\begin{equation}
	\Delta R_{xx} = R_0 - R_{xx}^{\mathrm{min}},
\end{equation}
and the normalized suppression is
\begin{equation}
	\frac{\Delta R_{xx}}{R_0}
	=
	\frac{R_0 - R_{xx}^{\mathrm{min}}}{R_0}.
\end{equation}

By construction, $\Delta R_{xx}/R_0 = 0$ corresponds to the absence of a resolved minimum, while $\Delta R_{xx}/R_0 = 1$ corresponds to a fully developed zero-resistance state. The procedure is applied independently to each $R_{xx}(\nu)$ trace at fixed $D$.

\section{Data on $\nu = 9/2$}

\subsection{Evolution of neighboring odd denominator FQHSs with perpendicular displacement field $D$ around $9/2$.}

\begin{figure*}[h]
	\includegraphics[width=1\columnwidth]{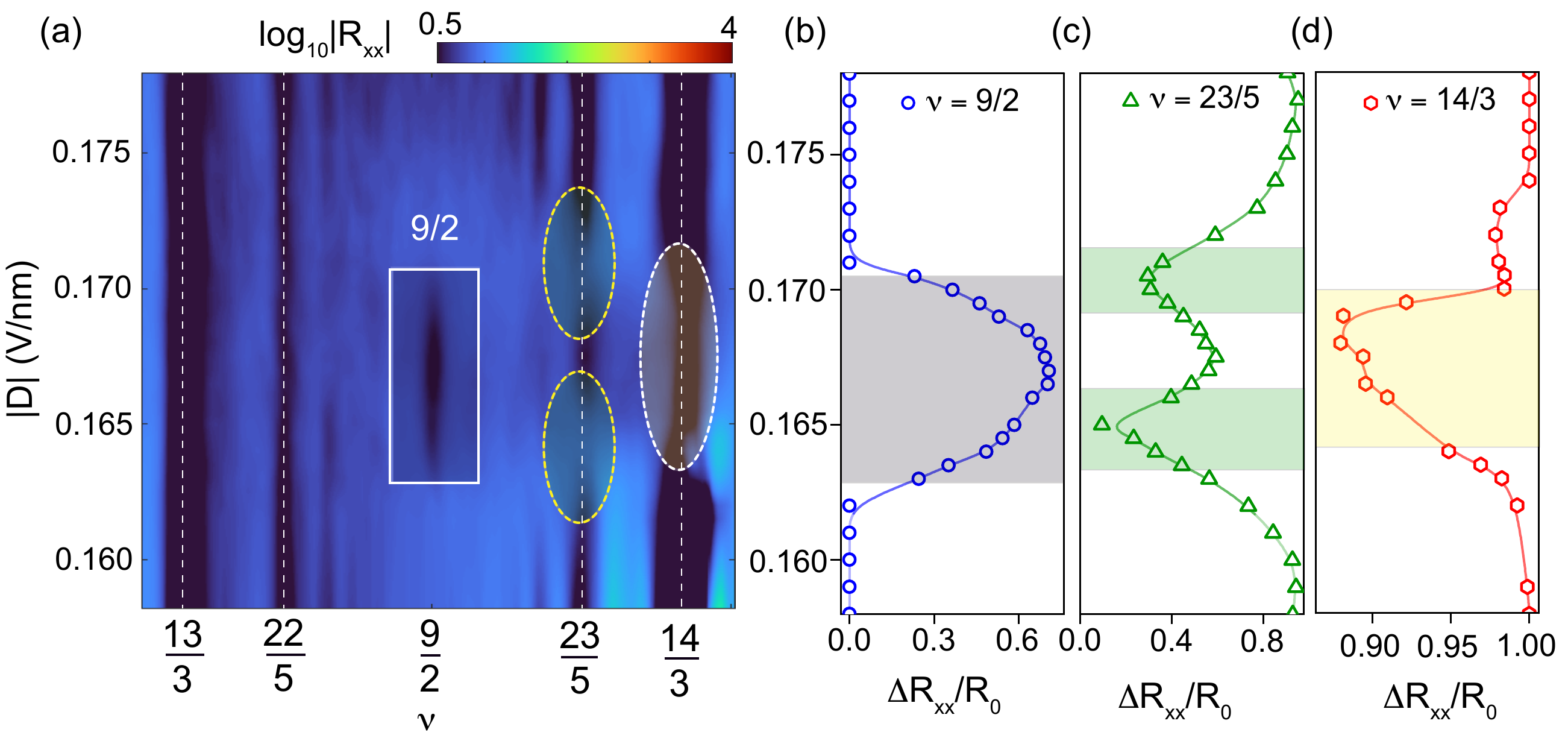}
	\caption{\textbf{Evolution of neighboring odd denominator states near $\nu=9/2$ (device 2).} (a) Color plot of $R_{xx}$ as a function of $\nu$ and $|D|$ in the vicinity of $\nu=9/2$ at $B=16$~T. The solid open white rectangle marks the $\nu=9/2$ state. Dashed lines mark the nearby odd denominator states. Plots of $\Delta R_{xx}/R_{0}$ versus $D$ for (b) $\nu=9/2$,(c) $\nu=23/5$,(d) $\nu=14/3$.}
	\label{fig:figS3}
\end{figure*}

The emergence of the even-denominator state at $\nu = 9/2$ in the $N=0$ LL of ABA-TLG is accompanied by systematic modifications of neighboring odd-denominator FQHSs. \ref{fig:figS3}(a) shows $R_{xx}$ as a function of $\nu$ and $|D|$ near $\nu = 9/2$ at $B=16$~T. The solid white rectangle marks the regime where $\nu = 9/2$ is maximally developed, while dashed lines indicate the position of odd-denominator states. To quantify the evolution of $\nu = 23/5$ and $\nu = 14/3$ states, we extract the normalized suppression $\Delta R_{xx}/R_0$ (Sec.~S2) versus $D$ for $\nu = 9/2$, $23/5$, and $14/3$ (\ref{fig:figS3}(b–d)).

The $\nu = 9/2$ state forms over a finite window $D \in [0.163,0.17]~\mathrm{V/nm}$, reaching maximal suppression $\Delta R_{xx}/R_0 \approx 0.7$ at $D = D_{9/2}^{\mathrm{max}} \approx 0.167~\mathrm{V/nm}$. The $\nu = 14/3$ state exhibits pronounced weakening precisely at $D_{14/3}^{\mathrm{min}} \approx 0.168~\mathrm{V/nm}$, coincident with $D_{9/2}^{\mathrm{max}}$. In contrast, $\nu = 23/5$ weakens only near the boundaries of the $\nu = 9/2$ window, at $D \approx 0.165~\mathrm{V/nm}$ and $0.17~\mathrm{V/nm}$, corresponding to the onset and termination of the paired phase, and remains comparatively stable away from the crossing.

This hierarchy demonstrates correlated competition between the even-denominator phase and neighboring composite-fermion states. The maximal suppression of $\nu = 14/3$ at optimal $D$, together with boundary-selective weakening of $\nu = 23/5$, indicates that the instability is localized near the LL crossing rather than reflecting a global change in interaction strength. The distinct $D$-dependence supports a crossing-driven pairing mechanism in which enhanced mixing of valley-resolved $N=0$ LLs destabilizes certain composite-fermion states while stabilizing the paired phase.

\subsection{Evolution of daughter states}

\begin{figure*}[h]
	\includegraphics[width=1\columnwidth]{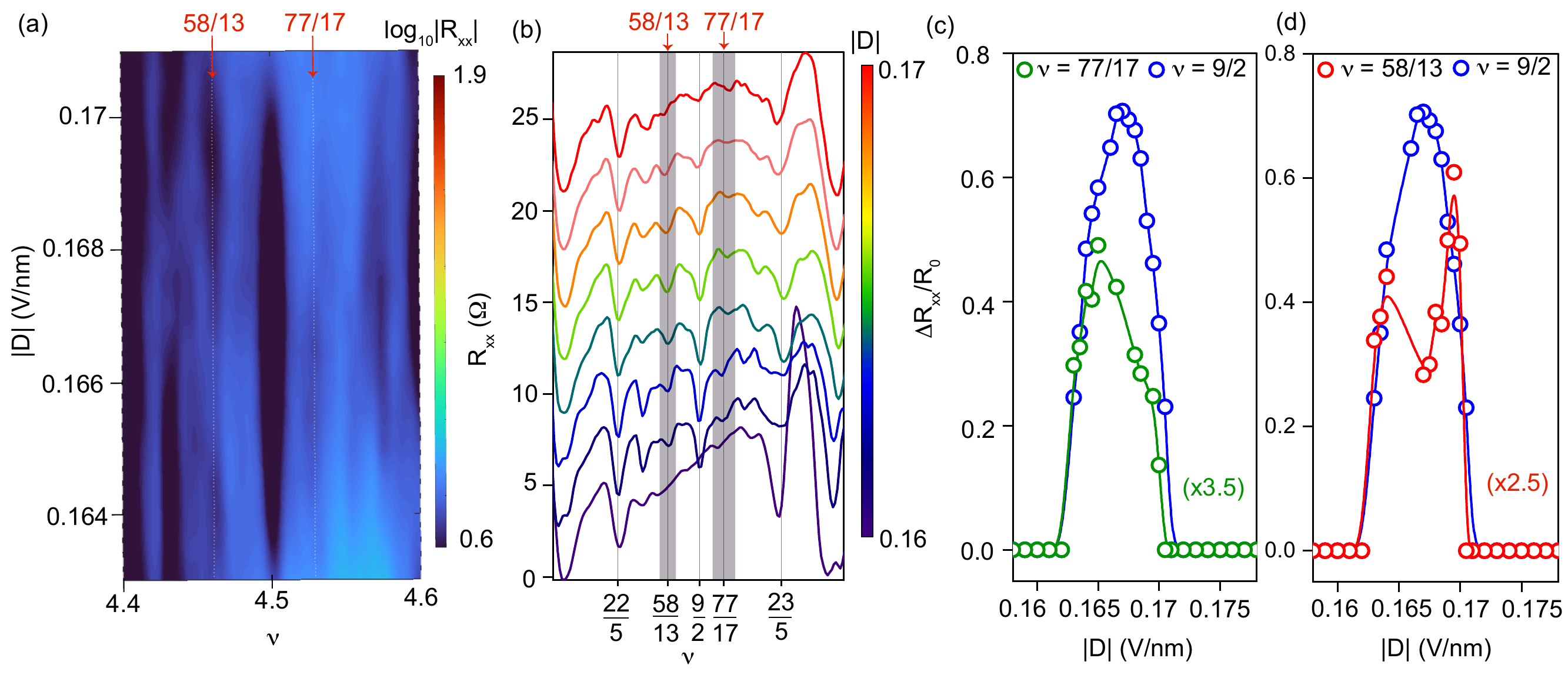}
	\caption{\textbf{Evolution of daughter states of $\nu=9/2$.} (a) Color plot of $R_{xx}$ as a function of $\nu$ and $|D|$ in the vicinity of $\nu=9/2$ at $B=16$~T. The daughter states are marked with red arrows. (b) Plot of $R_{xx}$ versus $\nu$ in the vicinity of $\nu=9/2$ at $B=16$~T. The plots are offset vertically by $3~\Omega$ for visibility. The shaded gray region marks the evolution of the daughter states. Plot of $\Delta R_{xx}/R_{0}$ as a function of $D$ for (c) $\nu=9/2~\mathrm{and}~77/17$, (d) $\nu=9/2~ \mathrm{and}~58/13$. Here, $\Delta R_{xx}/R_{0}$ of $\nu=77/17$ and $\nu=58/13$ are multiplied by constant factor to compare the evolution with $\nu=9/2$.}
	\label{fig:figS4}
\end{figure*}

\ref{fig:figS4} shows the displacement-field evolution of the $\nu=9/2$ daughter states at $B=16$~T. In \ref{fig:figS4}(a), minima in $R_{xx}$ at $\nu=58/13$, $9/2$, and $77/17$ appear in the $\nu$–$D$ color map. The daughter states are marked with red arrows. Line cuts $R_{xx}(\nu)$ at representative $D$ and at $B=16$~T are shown in \ref{fig:figS4}(b), where the shaded gray region marks the $\nu$ interval over which both daughters are resolved; outside this window, the minima vanish and the system becomes compressible.

The normalized suppression $\Delta R_{xx}/R_0$ (Sec.~S2) for $\nu=9/2$, $77/17$, and $58/13$ is plotted in \ref{fig:figS4}(c,d). The $\nu=77/17$ state closely tracks $\nu=9/2$, with both reaching maximal suppression at $D \approx D_{9/2}^{\mathrm{max}} \approx 0.167~\mathrm{V/nm}$ and weakening together outside the crossing window. In contrast, $\nu=58/13$ shows weaker, slightly asymmetric $D$-dependence, with shallow minima near $D \approx 0.167~\mathrm{V/nm}$. This asymmetry may arise from differences in excitation gap, disorder sensitivity, or interaction renormalization near the crossing.

The correlated $D$-dependence of both daughters with the parent $\nu=9/2$ FQHS demonstrates that the entire hierarchy emerges from the same crossing-enhanced instability. The common displacement-field window in which $\nu=9/2$, $77/17$, and $58/13$ are stabilized indicates a shared modification of the effective interaction near the valley-resolved $N=0$ LL crossing.

\subsection{$\nu=9/2$ state at $+D$ and $-D$.}

\begin{figure*}[h]
	\includegraphics[width=0.9\columnwidth]{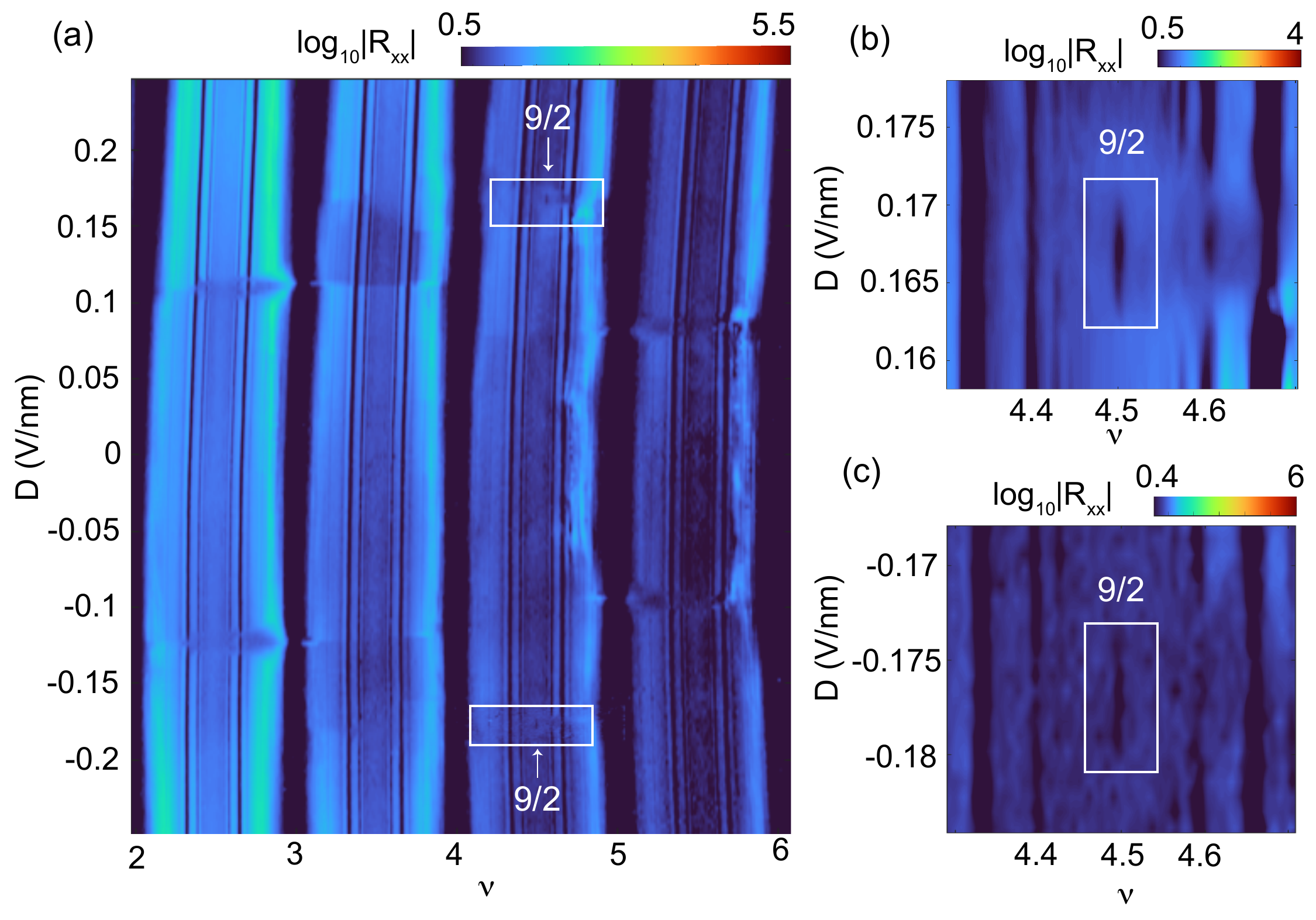}
	\caption{\textbf{$\nu = 9/2$ FQHS at $+D$ and $-D$} (a) Color plot of $R_{xx}$ as a function of $\nu$ and $D$ at $B=16$ T. Here, white open rectangle marks the region where $\nu=9/2$ forms. Zoomed in color plot of $R_{xx}$ in the vicinity of $\nu=9/2$ for (b) $+D$, (c) $-D$.}
	\label{fig:figS5}
\end{figure*}

\ref{fig:figS5}(a) shows $R_{xx}$ as a function of $\nu$ and $D$ at $B=16$~T. The displacement field $D$ is defined by the relative electrostatic potential between the graphite gates: positive $D$ corresponds to an electric field along $+z$ (bottom to top layer), and negative $D$ to $-z$ (\ref{fig:figS1}(b)).

The $\nu=9/2$ FQHS is observed for both $\pm D$ whenever the relevant $N=0$ LLs approach and cross (\ref{fig:figS5}(b,c)). Although the full LL spectrum is not strictly symmetric under $D\rightarrow -D$, likely due to lattice asymmetry and screening effects, the paired phase persists on both sides of the crossing.

\newpage

\section{Data on $\nu = 7/2$}

The displacement-field evolution near $\nu = 7/2$ mirrors that of $\nu = 9/2$, with daughter states tracking the even-denominator instability and competing odd states suppressed near the crossing.

\subsection{Evolution of neighboring odd denominator FQHSs with perpendicular displacement field $D$ around $7/2$}

\begin{figure*}[h]
	\includegraphics[width=1\columnwidth]{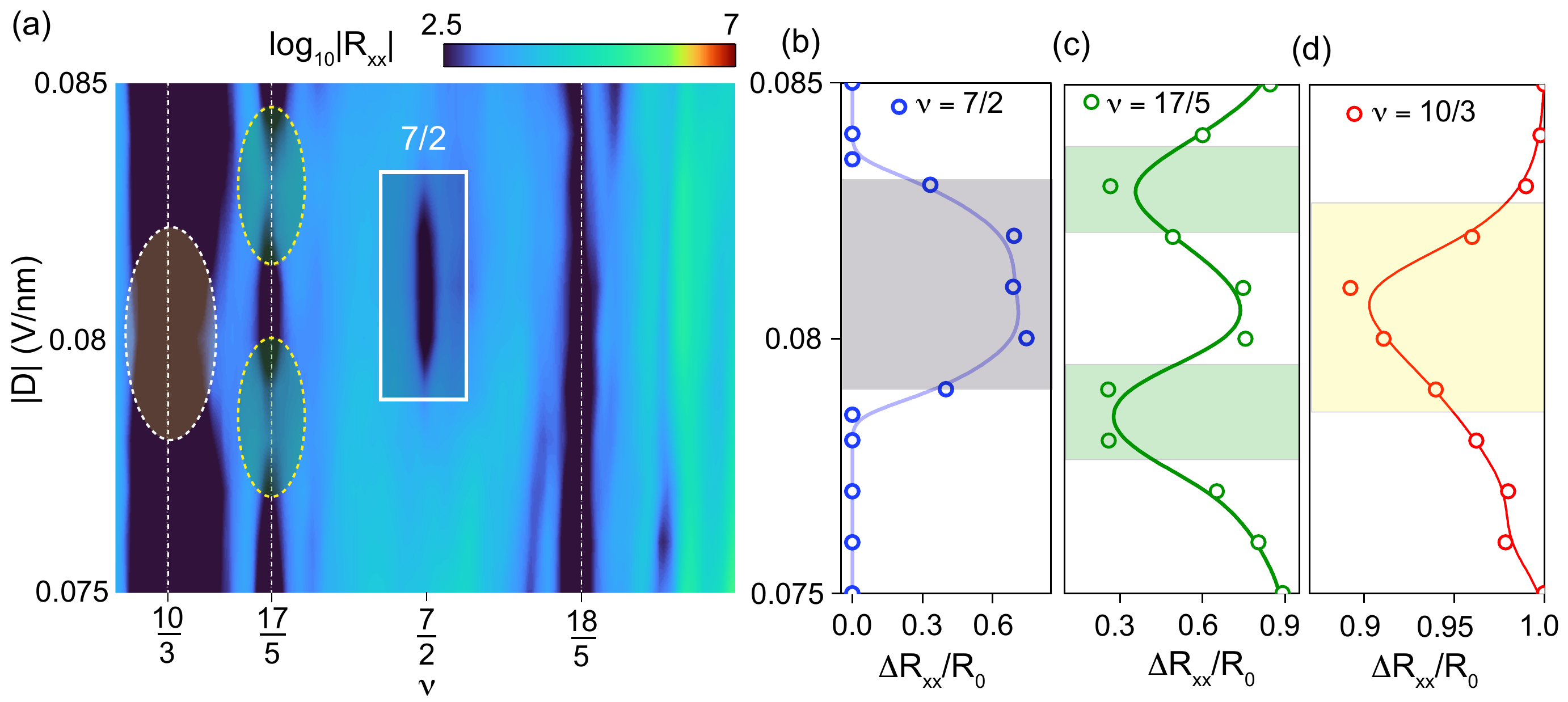}
	\caption{\textbf{Evolution of neighboring odd denominator states near $\nu=7/2$ (device 1).} (a) Color plot of $R_{xx}$ as a function of $\nu$ and $|D|$ in the vicinity of $\nu=7/2$ at $B=12$~T. The solid open white rectangle marks the $\nu=7/2$ state. Dashed lines mark the nearby odd denominator states. Dashed ellipses indicate the changes in the odd denominator states. Plot of $\Delta R_{xx}/R_{0}$ as a function of $D$ for (b) $\nu=7/2$,(c) $\nu=17/5$,(d) $\nu=10/3$.}
	\label{fig:figS6}
\end{figure*}

\ref{fig:figS6}(a) shows $R_{xx}$ as a function of $\nu$ and $|D|$ near $\nu=7/2$ at $B=12$~T. The solid white rectangle marks the regime where $\nu=7/2$ forms. The suppression of $\nu=10/3$ coincides with the displacement field at which $\nu=7/2$ is maximized, while $\nu=17/5$ weakens at two distinct values of $D$ ($0.083~\mathrm{V/nm}$ and $0.078~\mathrm{V/nm}$). This behavior is captured quantitatively in $\Delta R_{xx}/R_0$ for $\nu=7/2$ (\ref{fig:figS6}(b)), $\nu=17/5$ (\ref{fig:figS6}(c)), and $\nu=10/3$ (\ref{fig:figS6}(d)).

The maximal suppression of $\nu=10/3$ at optimal $D$, together with the double weakening of $\nu=17/5$ near the boundaries of the $\nu=7/2$ window, establishes correlated competition between the paired phase and neighboring composite-fermion states, consistent with a crossing-localized instability.

\subsection{Evolution of daughter states}

\begin{figure*}[h]
	\includegraphics[width=1\columnwidth]{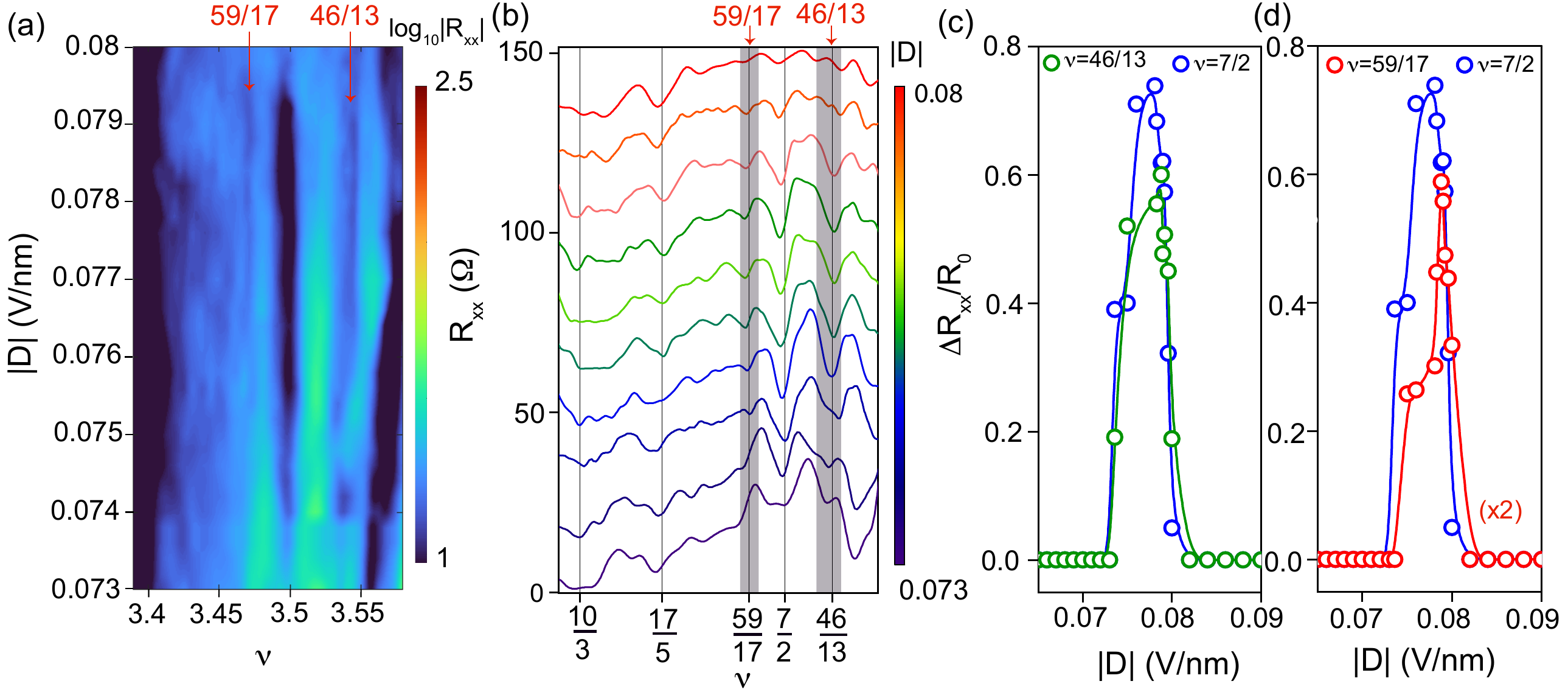}
	\caption{\textbf{Evolution of daughter states of $\nu=7/2$} (a) Color plot of $R_{xx}$ as a function of $\nu$ and $|D|$ in the vicinity of $\nu=7/2$ at $B=16$~T. Red arrows mark the daughter states. (b) Line plot of $R_{xx}$ as a function of filling factor $\nu$ in the vicinity of $\nu=7/2$ at different values of $D$ at $B=16T$. The plots are vertically offset by $14~\Omega$ factor for visibility. The shaded gray region marks the evolution of the daughter states. Plot of $\Delta R_{xx}/R_{0}$ as a function of $D$ for (c) $\nu=7/2,46/13$,(d) $\nu=7/2,59/17$. Here, $\Delta R_{xx}/R_{0}$ of $\nu=59/17$ is multiplied by constant factor to compare the evolution with $\nu=7/2$.}
	\label{fig:figS7}
\end{figure*}

\begin{figure*}[h]
	\includegraphics[width=0.9\columnwidth]{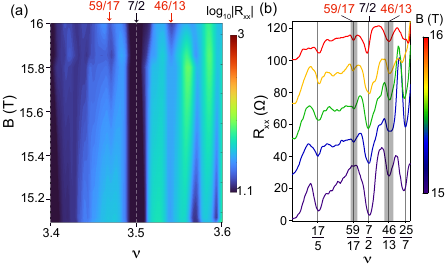}
	\caption{\textbf{Evolution of $\nu=7/2$ and its daughter states with $B$.} (a) Color plot of $R_{xx}$ as a function of $B$ and $\nu$ in the vicinity of $\nu=7/2$ at $D=-0.0785$~V/nm. (b) The corresponding line scans of $R_{xx}$ as a function of $\nu$. The shaded gray region marks the evolution of daughter states.  $R_{xx}$ curves are scaled vertical offset by a constant factor of $25$~$\Omega$ for visibility. \label{fig:figS8}}
\end{figure*}

\ref{fig:figS7} shows the displacement-field evolution of the $\nu=7/2$ daughter hierarchy at $B=16$~T. In \ref{fig:figS7}(a) the minima at  $\nu=59/17$ and $46/13$ are marked by red arrows. Line cuts in \ref{fig:figS7}(b) resolve their evolution, with the shaded region indicating the $\nu$ window over which the daughter states form.

The normalized suppression $\Delta R_{xx}/R_0$ for $\nu=7/2$, $46/13$, and $59/17$ shows that both daughters closely track the parent $\nu=7/2$ state, peaking within the same displacement-field window. For comparison, $\Delta R_{xx}/R_0$ of $\nu=59/17$ is scaled by a constant factor in \ref{fig:figS7}(c,d). The correlated $D$-dependence establishes that the daughter hierarchy arises from the same crossing-enhanced instability.

\ref{fig:figS8}(a) shows the evolution of $\nu=7/2$ and its daughters with magnetic field $B$ at constant displacement field $D=-0.0785$~V/nm. Corresponding $R_{xx}$ line cuts are shown in \ref{fig:figS8}(b). At $B=15$~T, $\nu=46/13$ is well developed, whereas $\nu=59/17$ emerges only once $B$ reaches $16$~T, indicating increasing stabilization of the daughter hierarchy with $B$-field.
\clearpage
\subsection{Disorder induced broadening and formation of $7/2$ state in narrow $D$-range.}

\begin{figure*}[h]
	\includegraphics[width=0.8\columnwidth]{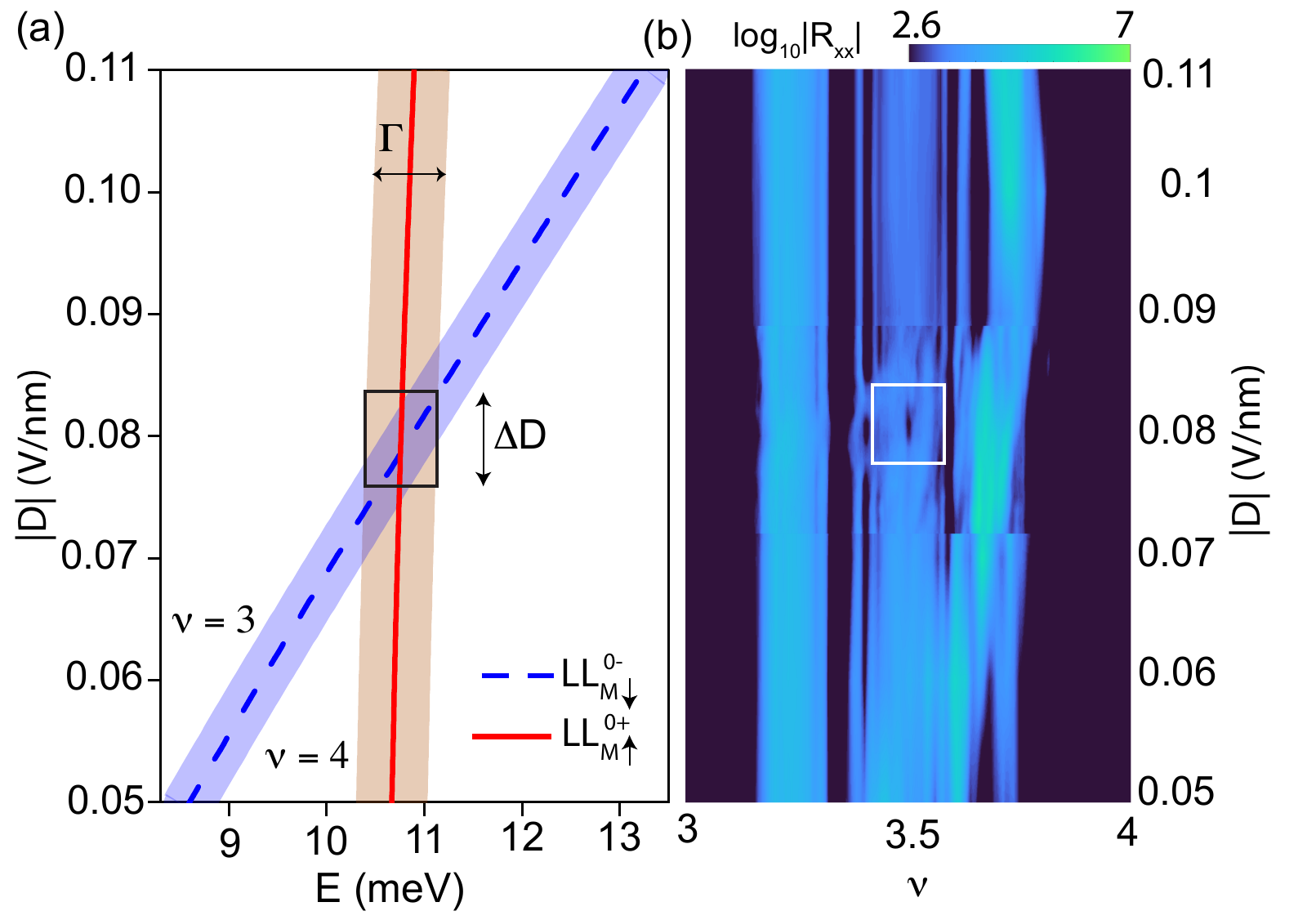}
	\caption{\textbf{Disorder induced LL broadening.} (a) Simulated LL spectrum versus  $|D|$ and $E$ at $B=12$~T. The shading marks the disorder-induced broadening of the LLs. The rectangle marks the region of overlap between the two disorder-broadened LLs. (b) Color plot of $R_{xx}$ as a function of $\nu$ and $|D|$ measured at $B=12$~T. The white solid rectangle marks the range of $D$ where $\nu=7/2$ is formed.}
	\label{fig:figS9}
\end{figure*}

The finite displacement-field window of the $\nu=7/2$ state can arise from disorder-induced LL broadening. In the absence of disorder, the crossing would occur at a single $D$; with spectral width $\Gamma$, LL overlap — and hence mixing — occurs over $\Delta D \sim \Gamma/(\partial E/\partial D)$. For the $\nu=7/2$ crossing, we estimate $\Gamma \approx 4$~K from Shubnikov–de Haas oscillations~\cite{frpm-3fs8}, yielding $\Delta D = 0.0065$~V/nm, consistent with experiment (\ref{fig:figS9}). The widths for different LL crossings are comparable but not identical due to variations in $\Gamma$ and in $\partial E/\partial D$.

\clearpage

\section{Data on $\nu=5/2$.}

\begin{figure*}[h]
	
	\includegraphics[width=\columnwidth]{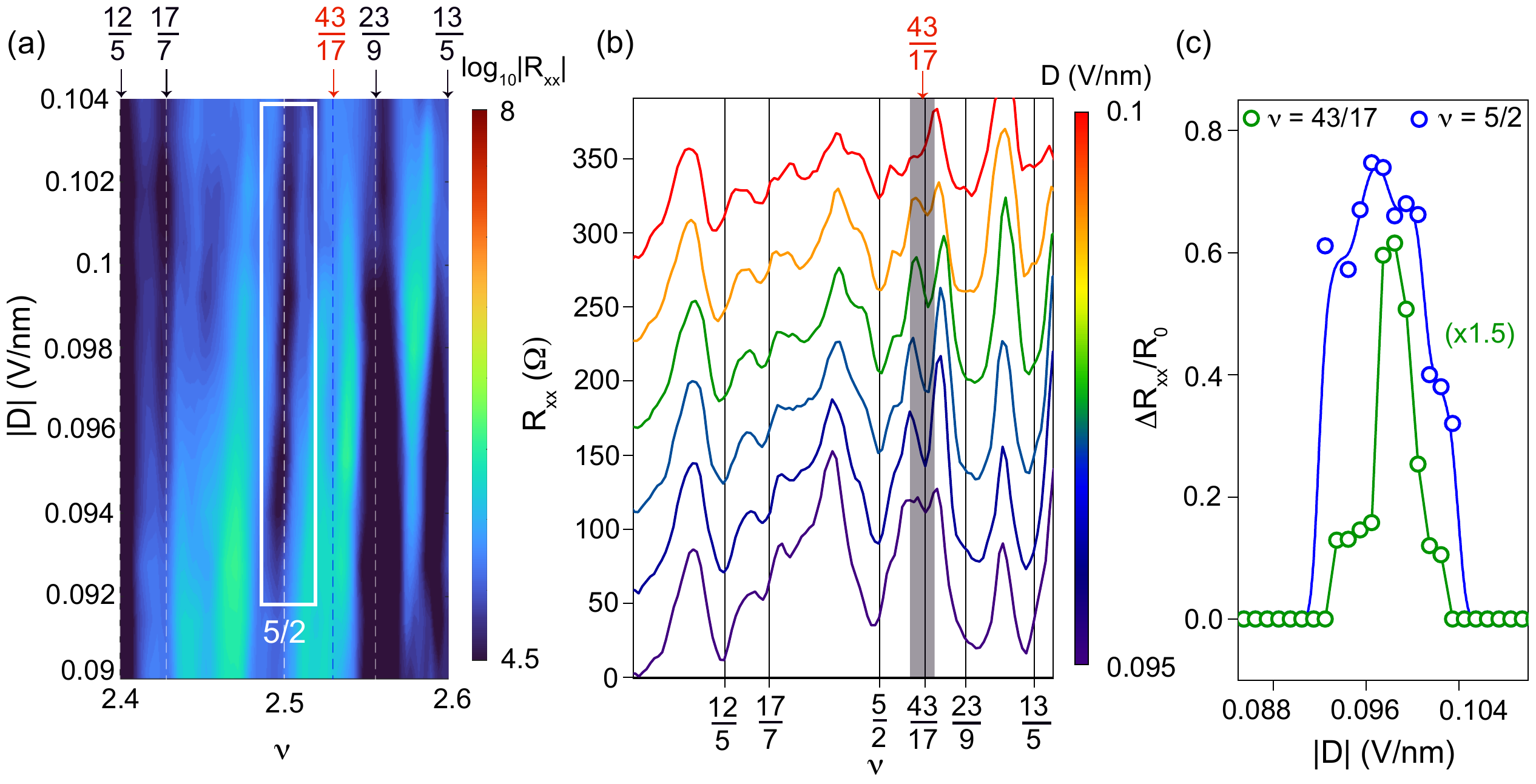}
	\caption{\textbf{Evolution of $\nu=5/2$ with displacement field (device 1).} (a) $R_{xx}$ as a function of $\nu$ and $D$ at $B=12$~T; the white rectangle marks the $\nu=5/2$ regime and the red arrow indicates $\nu=43/17$. (b) $R_{xx}$ versus $\nu$ at different $D$ and at $B=12$~T (offset by $57~\Omega$); the gray region highlights the daughter state. (c) $\Delta R_{xx}/R_{0}$ versus $D$ for $\nu=5/2$ and $\nu=43/17$.}
	\label{fig:figS11}
\end{figure*}

\ref{fig:figS11}(a) shows $R_{xx}$ as a function of $\nu$ and $D$ measured at $B=12$~T. The white rectangle marks the displacement-field window in which $\nu=5/2$ forms, and the red arrow indicates its daughter state at $\nu=43/17$. Comparison with the simulated spectrum in Fig.~4(a) of the main text shows that the $\nu=5/2$ FQHS emerges near the crossing of $LL_{M, \uparrow}^{0-}$ and $LL_{M,\uparrow}^{0+}$ LLs. Near $|D|=0.1$~V/nm, the activation gap of $\nu=5/2$ state is about $0.1$~K, while the maximum gaps of the $\nu=9/2$ and $7/2$ states reach about $0.5$~K. 

The displacement-field evolution is further resolved in the line cuts of $R_{xx}(\nu)$ shown in \ref{fig:figS11}(b), where the shaded region highlights the $\nu$ range over which the daughter state develops. 

The normalized suppression $\Delta R_{xx}/R_0$ for $\nu=5/2$ and $\nu=43/17$ is plotted in \ref{fig:figS11}(c). The minima of the daughter state closely track the evolution of the parent $\nu=5/2$ state, demonstrating correlated stabilization within the same crossing window. This correspondence confirms that the $\nu=43/17$ state originates from the same crossing-enhanced instability responsible for the $\nu=5/2$ FQHS.

\section{Energy gap of the $4^{\mathrm{th}}$ LL}

\begin{figure*}[h]
	\includegraphics[width=0.9\columnwidth]{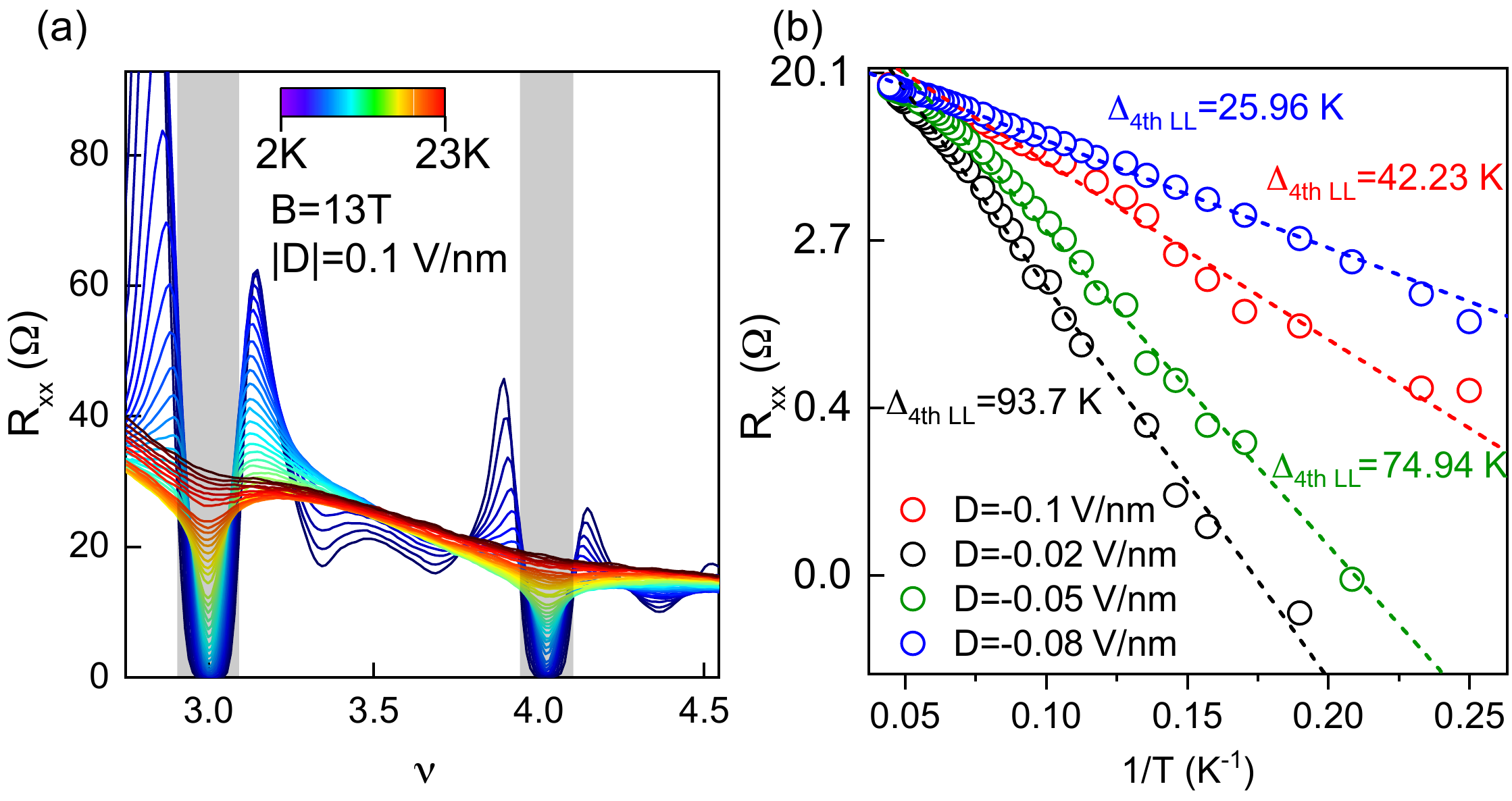}
	\caption{\textbf{Activation gaps of the fourth LL.} (a) Line plots of $R_{xx}$ as a function of  $\nu$ at several representative temperatures in the range $2$~K$<T<$$23$~K, measured at $B=13$~T and $|D|=0.1$~V/nm. (b) Arrhenius fits to $R_{xx}$ at representative values of $D$. }

	\label{fig:figS13}
\end{figure*}

\ref{fig:figS13}(a) shows $R_{xx}$ as a function of $\nu$ at different temperatures measured at $B=13$~T and $|D|=0.1$~V/nm near $LL=4$. Activation gaps are extracted from Arrhenius fits to the temperature dependence of the $R_{xx}$ minima at various $D$, shown in \ref{fig:figS13}(b). The fits follow the relation
\begin{equation}
	R_{xx} = R_0 \exp\left(-\frac{\Delta}{2k_B T}\right),
	\label{Eqn:arrehniuseq}
\end{equation}
from which $\Delta$ is obtained from the slope.

The extracted gaps as a function of $D$ are plotted in Fig.~4(b,c) of the main manuscript.

\section{Details of simulations}

The Landau spectrum is calculated using the Slonczewski–Weiss–McClure tight-binding parametrization for ABA-TLG~\cite{Abanin13, PhysRevLett.117.066601, Chen2024}. The model contains six hopping parameters $\{\gamma_0,\gamma_1,\dots,\gamma_5\}$ corresponding to
\begin{subequations}
	\begin{align}
		A_i \leftrightarrow B_i &: \gamma_0, \quad
		B_{1/3} \leftrightarrow A_2 : \gamma_1 \\
		A_1 \leftrightarrow A_3 &: \gamma_2/2, \quad
		A_{1/3} \leftrightarrow B_2 : \gamma_3 \\
		(A/B)_{1/3} \leftrightarrow (A/B)_2 &: -\gamma_4, \quad
		B_1 \leftrightarrow B_3 : \gamma_5/2,
	\end{align}
\end{subequations}
where $A(B)$ denotes the sublattice and $i=1\dots3$ labels the layer. An additional onsite parameter $\delta$ accounts for the vertical dimer sites ($B_1$, $B_3$, $A_2$).

In the basis $\{A_1,B_1,A_2,B_2,A_3,B_3\}$, the Hamiltonian is
\begin{equation}
	H_0 =
	\begin{pmatrix}
		0 & \gamma_0 t_\mathbf{k}^* & \gamma_4 t_\mathbf{k}^* & \gamma_3 t_\mathbf{k} & \gamma_2/2 & 0 \\
		\gamma_0 t_\mathbf{k} & \delta & \gamma_1 & \gamma_4 t_\mathbf{k}^* & 0 & \gamma_5/2 \\
		\gamma_4 t_\mathbf{k} & \gamma_1 & \delta & \gamma_0 t_\mathbf{k}^* & \gamma_4 t_\mathbf{k} & \gamma_1 \\
		\gamma_3 t_\mathbf{k}^* & \gamma_4 t_\mathbf{k} & \gamma_0 t_\mathbf{k} & 0 & \gamma_3 t_\mathbf{k}^* & \gamma_4 t_\mathbf{k} \\
		\gamma_2/2 & 0 & \gamma_4 t_\mathbf{k}^* & \gamma_3 t_\mathbf{k} & 0 & \gamma_0 t_\mathbf{k}^* \\
		0 & \gamma_5/2 & \gamma_1 & \gamma_4 t_\mathbf{k}^* & \gamma_0 t_\mathbf{k} & \delta
	\end{pmatrix},
\end{equation}
with $t_\mathbf{k} = \sum_{j=1}^3 e^{i\mathbf{k}\cdot\mathbf{a}_j}$, lattice vectors $\mathbf{a}_0=a(0,1/\sqrt3)$ and $\mathbf{a}_{1/2}=a(\mp 1/2,-1/2\sqrt3)$, and $a=2.46$~\AA.

Expanding near $K^\pm$ and substituting $\gamma_i t_\mathbf{k}\rightarrow v_i\pi$ yields the low-energy Hamiltonian, where $\pi=\xi k_x+i k_y$ and $\hbar v_i=\frac{\sqrt3}{2}a\gamma_i$.

\begin{figure}
	\centering
	\includegraphics[width=0.75\linewidth]{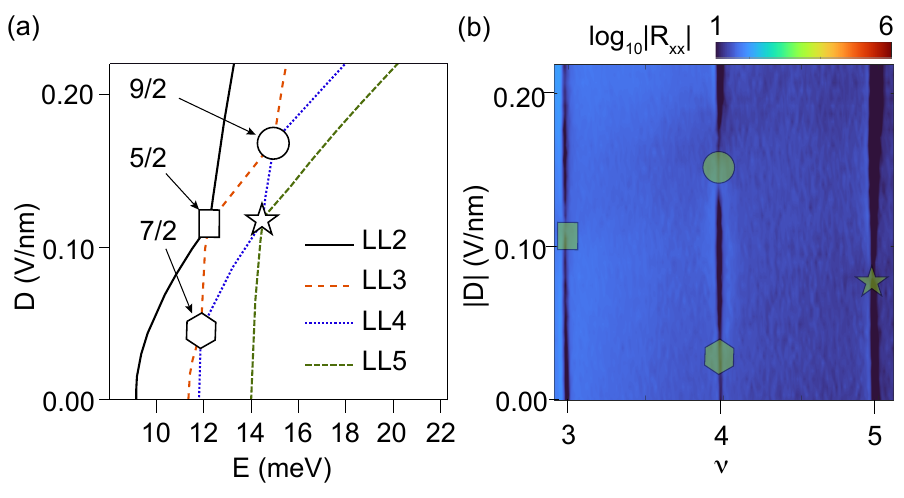}
	\caption{\textbf{Crossing of four $N=0$ LLs in ABA-TLG.} 
		(a) Simulated monolayer-like zeroth LL spectrum as a function of displacement field $D$ and energy at $B=16$~T. The four spin- and valley-resolved $N=0$ LLs (LL2–LL5) are shown, and symbols mark the calculated crossing points corresponding to $\nu=7/2$, $5/2$, and $9/2$. 
		(b) Measured $\log_{10}|R_{xx}|$ as a function of filling factor $\nu$ and $|D|$ at $B=16$~T and $T=8.5$~K. The weakening of adjacent integer quantum Hall states at the symbol-marked $D$ values reflects enhanced LL mixing near the crossings.}
	\label{fig:figS14}
\end{figure}

A perpendicular electric field is incorporated via layer potentials $V_1\dots V_3$, parametrized by
\begin{equation}
	\Delta_1 = (-e)\frac{V_1-V_2}{2}, \qquad
	\Delta_2 = (-e)\frac{V_1+V_3-2V_2}{6}.
\end{equation}
In the absence of an external field, transforming to the basis $\{\frac{A_1-A_3}{\sqrt2},\frac{B_1-B_3}{\sqrt2},\frac{A_1+A_3}{\sqrt2},B_2,A_2,\frac{B_1+B_3}{\sqrt2}\}$ block-diagonalizes the Hamiltonian into monolayer-like and bilayer-like sectors,
\begin{equation}
	H_0+H_{\Delta_2}=
	\begin{pmatrix}
		H_{slg} & 0\\
		0 & H_{blg}
	\end{pmatrix},
\end{equation}
while the external field couples the blocks through
\begin{equation}
	H_{\Delta_1}=
	\begin{pmatrix}
		0 & H_{ext}\\
		H_{ext} & 0
	\end{pmatrix}, \qquad
	H_{ext}=
	\begin{pmatrix}
		\Delta_1 & 0 & 0 & 0\\
		0 & 0 & 0 & \Delta_1
	\end{pmatrix}.
\end{equation}

The monolayer-like block is
\begin{equation}
	H_{slg}=
	\begin{pmatrix}
		\Delta_2-\gamma_2/2 & v_0\pi^\dagger\\
		v_0\pi & -\gamma_5/2+\delta+\Delta_2
	\end{pmatrix},
\end{equation}
and the bilayer-like block is
\begin{equation}
	H_{blg}=
	\begin{pmatrix}
		\gamma_2/2+\delta & \sqrt2 v_3\pi & -\sqrt2 v_4\pi^\dagger & v_0\pi^\dagger\\
		\sqrt2 v_3\pi^\dagger & -2\Delta_2 & v_0\pi & -\sqrt2 v_4\pi\\
		-\sqrt2 v_4\pi & v_0\pi^\dagger & \delta-2\Delta_2 & \sqrt2\gamma_1\\
		v_0\pi & -\sqrt2 v_4\pi^\dagger & \sqrt2\gamma_1 & \gamma_5/2+\delta+\Delta_2
	\end{pmatrix}.
\end{equation}

The parameters used are $\gamma_0=3.1$~eV, $\gamma_1=0.39$~eV, $\gamma_2=-0.005$~eV, $\gamma_3=0.275$~eV, $\gamma_4=0.040$~eV, $\gamma_5=0.005$~eV, $\delta=0.0108$~eV, and $\Delta_2=0.003$~eV.

Landau quantization is introduced via $\pi\rightarrow\pi-e(A_x+iA_y)$ and Landau gauge $A_x=0$, $A_y=Bx$. With conserved $k_y$, 
\begin{equation}
	\pi=\frac{-i\hbar}{l_B}(\xi\partial_x+x-k_y).
\end{equation}
In the LL basis,
\begin{subequations}
	\begin{align}
		K^+:\;\pi|n\rangle &= \frac{i\hbar}{l_B}\sqrt{2(n+1)}|n+1\rangle, \\
		K^+:\;\pi^\dagger|n\rangle &= -\frac{i\hbar}{l_B}\sqrt{2n}|n-1\rangle, \\
		K^-:\;\pi|n\rangle &= \frac{i\hbar}{l_B}\sqrt{2n}|n-1\rangle, \\
		K^-:\;\pi^\dagger|n\rangle &= -\frac{i\hbar}{l_B}\sqrt{2(n+1)}|n+1\rangle.
	\end{align}
\end{subequations}

A finite LL cutoff $\alpha$ is imposed, replacing $\pi$ by $\alpha\times\alpha$ matrices. Spurious low-energy eigenvalues arising from truncation are removed by filtering states with significant $|n\rangle$ weight. We use $\alpha=100$ throughout.

\section{Role of inversion symmetry in valley structure.}

\begin{figure}[h]
	\includegraphics[width=0.5\columnwidth]{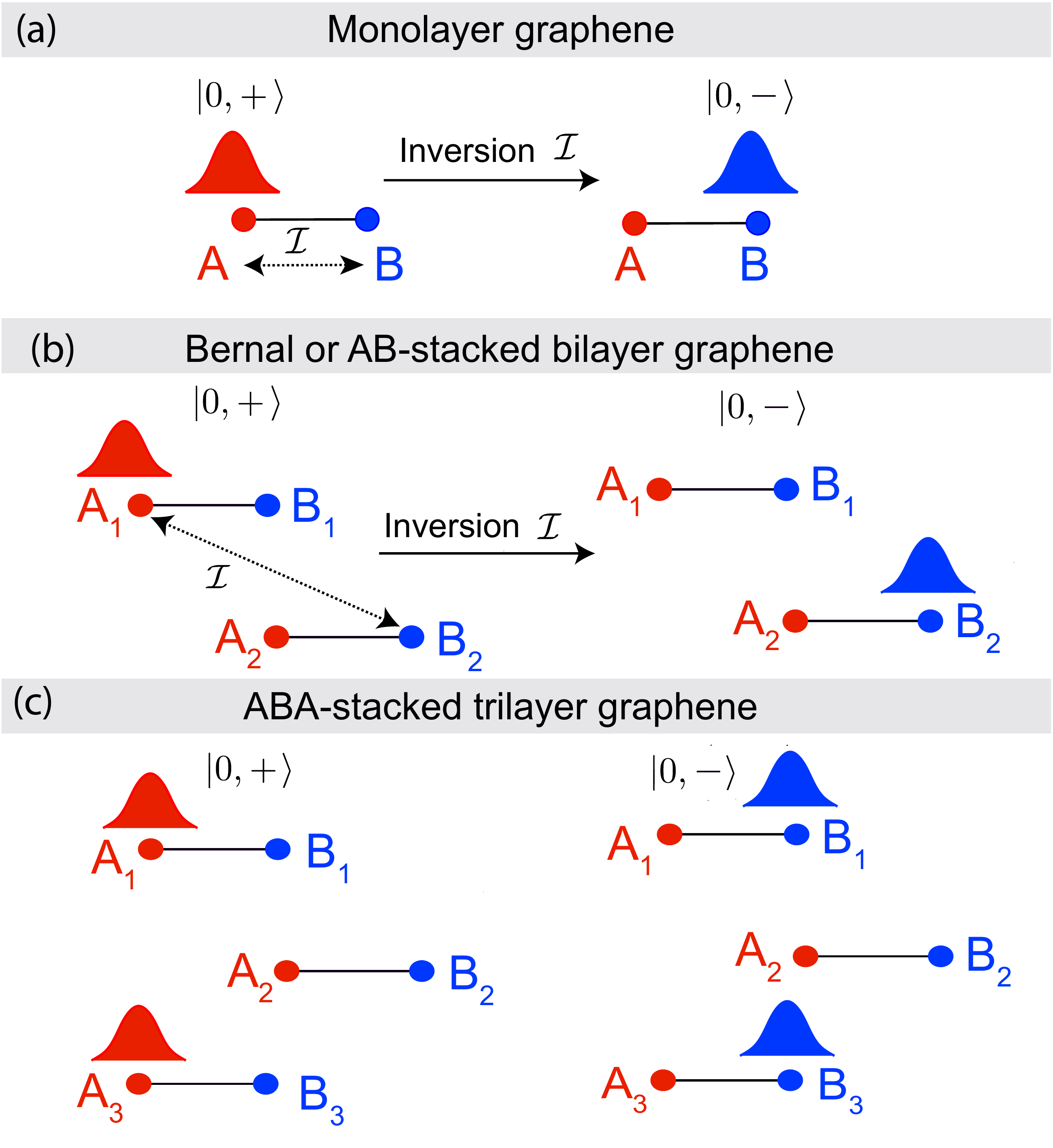}
	\small{\caption{\textbf{Role of inversion symmetry in valley structure.} (a,b) Schematic $N = 0$ LL wavefunctions in MLG and BLG, which are invariant under inversion $\mathcal{I}$ and preserve valley equivalence.  (c) Zeroth LL wavefunctions in ABA-TLG. The absence of inversion symmetry (as evidenced by distinct local charge environments) renders the $|0,+\rangle$ and $|0,-\rangle$ states inequivalent, enabling valley-dependent interaction renormalization and intra-orbital mixing near LL crossings.
			\label{fig:fig5}}}
\end{figure}

In MLG and BLG, inversion symmetry renders the two valleys equivalent and preserves the valley symmetry of the $N=0$ LL wavefunctions (Fig.~\ref{fig:fig5}(a,b)), suppressing intra-orbital valley hybridization; substrate-induced valley splitting in MLG does not alter the underlying lattice inversion symmetry. By contrast, ABA-TLG lacks inversion symmetry, making the $|0,+\rangle$ and $|0,-\rangle$ states inequivalent (Fig.~\ref{fig:fig5}(c)) and permitting valley-dependent interaction renormalization near a LL crossing.     

    \clearpage
    \bibliography{arxiv}

\end{document}